# Neuronal Synchronization Can Control the Energy Efficiency of Inter-Spike Interval Coding

Siavash Ghavami[1*#], Vahid Rahmati[2,3*], Farshad Lahouti[4], Lars Schwabe[5]

[1]Department of Electrical and Computer Engineering, University of Minnesota, Minneapolis, MN, 55455,

[3] Department of Psychology, Technical University of Dresden, Dresden, Germany.

[4] Hans-Berger Department of Neurology, University Hospital Jena, Jena, Germany.

[5] Department of Electrical Engineering, California Institute of Technology, Pasadena, CA, 91125, USA.

[6] Lufthansa Industry Solutions, Data Insight Lab, Schützenwall 1, D-22844 Norderstedt, Germany

* Authors contributed equally in this paper.

# Correspondence: siavash.ghavami@gmail.com

ABSTRACT

The role of synchronous firing in sensory coding and cognition remains controversial. While studies, focusing on its mechanistic consequences in attentional tasks, suggest that synchronization dynamically boosts sensory processing, others failed to find significant synchronization levels in such tasks. We attempt to understand both lines of evidence within a coherent theoretical framework. We conceptualize synchronization as an independent control parameter to study how the postsynaptic neuron transmits the average firing activity of a presynaptic population, in the presence of synchronization. We apply the Berger-Levy theory of energy efficient information transmission to interpret simulations of a Hodgkin-Huxley-type postsynaptic neuron model, where we varied the firing rate and synchronization level in the presynaptic population independently. We find that for a fixed presynaptic firing rate the simulated postsynaptic interspike interval distribution depends on the synchronization level and is well-described by a generalized extreme value distribution. For synchronization levels of 15% to 50%, we find that the optimal distribution of presynaptic firing rate, maximizing the mutual information per unit cost, is maximized at ~30% synchronization level. These results suggest that the statistics and energy efficiency of neuronal communication channels, through which the input rate is communicated, can be dynamically adapted by the synchronization level.

I. INTRODUCTION

Selective attention affects early stages of sensory processing [1] but the detailed underlying mechanisms are not fully understood. One theory proposes that neural activity representing the attended stimuli is selected for further processing via synchronization [2]. While this is supported by some recent studies [3, 4], other studies fail to find significant effects [5]. Modeling studies [6, 7] have been built upon the idea that synchronous firing affects the firing rate [8, 9] and could dynamically modulate the signal flow [10].

Neuro-spike communication has been recently widely studied from information theory and communication theory perspectives [11-16]. These include studies of multi-terminal neuro-spike communication [11, 17], error probability of neuro-spike communication [13], effects of temporal modulation and spike shape variations on capacity of nurospike communication [16], and effect of randomness in capacity release [16].

The principled framework of information theory [18] has been successfully applied to study early sensory coding [19-22], and the mutual information (MI) has been used as a quantitative measure of information content of neuronal responses. For example, it was found that synchronous thalamic discharges carry contrast-invariant information about the orientation of visual stimuli [23]. This suggests that synchronization itself may be the information-carrying signal as opposed to only it being the signature of a mechanism that modulates the signal flow. Here we ask the question "What could be the role of synchronization in transmitting information?", and address it using an energy aware information theoretic framework. We studied this framework partially in [24].

We adopt the notion that synchronization modulates signal flow [10] and consider the synchronization within a presynaptic population as an independent control parameter adjusting the "channel characteristics" of postsynaptic neurons, i.e., we conceptualize them as dynamically configurable communication channels [25]. The Berger-Levy theory goes beyond MI maximization by postulating the maximization of capacity per unit cost (measured in bits per joule, bpj) as the objective in neuronal communication [24, 26, 27]. In that line, energy-efficiency has already been suggested for retina [28, 29] and cortex [30-32], but normative modeling within the Berger-Levy theory remains rare [32-36].

In computer simulations we vary independently the presynaptic firing rate (the input) and synchronization level and characterize the transformation into postsynaptic interspike intervals (ISI; the output) in terms of the conditional probability P(*Output* | *Input*; *Synchronization level*). We find that i) the conditional probability is well-described by a generalized extreme value (GEV) distribution. For synchronization levels of 15% to 50% (restricted for technical reasons) we ii) compute the optimal P(*Input*) that maximizes the MI per unit cost. Most interestingly, we find that iii) the MI per unit cost has a maximum at approximately 30% synchronization level. These findings suggest that attentional modulations, acting via changing the synchronization level, could be conceptualized as modulating dynamically the energy-efficiency of information flow in a potentially task- and context-dependent manner.

The remainder of this paper is organized as follows. In section II the methods are presented which includes: system and neuron model, modeling the presynaptic inputs, the neuronal synchronization, the balanced regime, parametric distributions, statement of the optimization problem. In section III results are presented which includes: the conditional distribution of postsynaptic ISI, optimal input rate distributions, dependency of the energy efficiency on the synchronization level. In section IV discussions are presented which studies: effect of neuronal synchronization on the information transmission during attention, the role of neuronal synchronization in modulating the firing characteristics of postsynaptic neuron, the implications of neuronal synchronization for energy efficiency of neuronal coding, extensions and limitations.

## II. METHODS

We simulated a cortical neuron which receives balanced excitatory and inhibitory synaptic inputs. The neuron model is a one-compartment Hodgkin-Huxley-type (HH-type) model with an adaptation current, and the synapses modeled as non-saturating instantaneous conductance changes with exponential decay. Presynaptic spike trains were generated using independent Poisson processes into which we injected synchronous spiking events to control the synchronization level of presynaptic neurons, while keeping the average firing rate, as sensed at each single synapse, constant. For different synchronization levels we simulated the response of the postsynaptic neuron model, estimated the histogram of postsynaptic ISI for different input rates, and fitted these histograms with parametric distributions. Then, an optimization problem was defined to find an optimal level of synchronization, where optimality is defined in terms of maximizing the capacity per unit cost for the neuronal communication via a postsynaptic ISI code. In the rest of this Methods section we describe in greater detail the system model, the neuron model, and the synaptic inputs model for two simulation scenarios. Finally, we state the optimization problem.

### A. System Model

We model a neuron as a *communication channel* (see Figure 1). The *channel inputs* are mean firing rates of presynaptic excitatory neurons, $\lambda_{exc}$. The *channel output* is the ISI of the postsynaptic neuron. The *channel characteristics* are given by the ISI distribution of the postsynaptic neuron that depends on various factors, including not only the $\lambda_{exc}$, but also the mean firing rates of presynaptic inhibitory neurons, $\lambda_{inh}$, and the synchronization level $s$ among the presynaptic excitatory neurons. We consider the synchronization level as a control parameter of the channel that affects its characteristics. We use the notation $f_T(\tau|\lambda_{exc},\lambda_{inh},s)$ to denote the corresponding conditional distribution of postsynaptic ISI ($\tau$) with given $\lambda_{exc}$ as the input of channel.

We obtain the postsynaptic ISI distributions by simulating a HH-type neuron model using two specific simulation scenarios motivated by two biophysically different transmission regimes, namely the excitation-driven and the fluctuation-driven. Both these scenarios emulate the so-called balanced regimes of excitatory and inhibitory synaptic inputs (see below). In the excitation-driven scenario we keep the inhibitory rate $\lambda_{inh}$ constant. In contrast, in the fluctuation-driven scenario we co-vary the inhibitory rate $\lambda_{inh}$ together with the excitatory rate $\lambda_{exc}$. In terms of the system model, in the excitation-driven scenario the constant inhibition corresponds to considering $\lambda_{inh}$ as a constant control parameter, whereas in the fluctuation-driven scenario the inhibition $\lambda_{inh}$ is a varying control parameter as function of the channel input, $\lambda_{exc}$. In both scenarios, however, the synchronization level $s$ among the excitatory neurons is the channel's control parameter we are primarily interested in. For brevity in the sequel, we drop $\lambda_{inh}$ in $f_T(\tau|\lambda_{exc}, \lambda_{inh}, s)$ and denote it with $f_T(\tau|\lambda_{exc}, s)$.

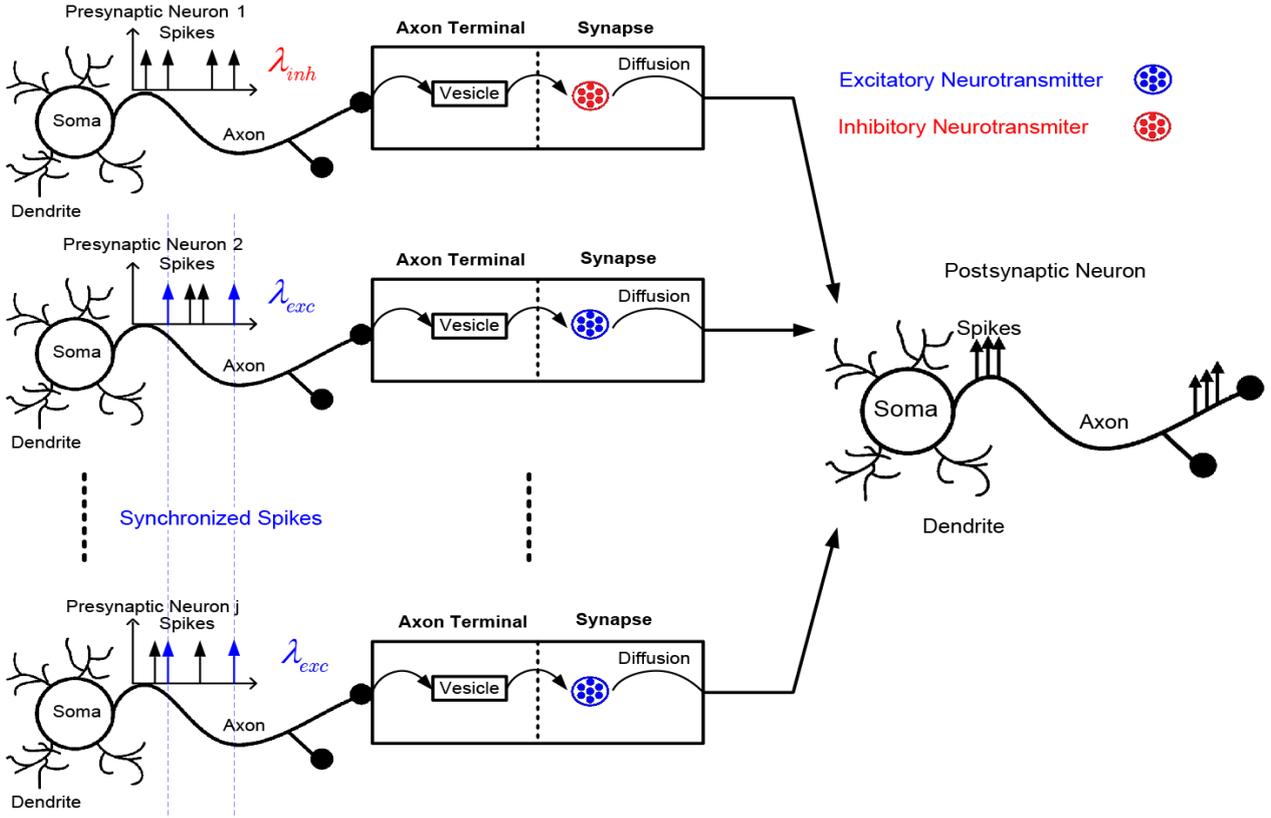

**Figure 1**. Illustration of the communication channel model. The excitatory and inhibitory neurons in the presynaptic population are firing spikes with mean rates $\lambda_{exc}$ and $\lambda_{inh}$, respectively. These rates are sent through the channel (the set of synapses projected onto the postsynaptic neuron) and are coded as postsynaptic ISIs. Some spikes of the excitatory neurons are synchronized (blue arrows). For each presynaptic neuron, the combination of the synchronous and asynchronous spikes determine its $\lambda_{exc}$. For different levels of synchronization, and potentially different levels of inhibition, the channel itself changes its characteristics as reflected by different conditional distributions of postsynaptic ISI, i.e., $f_T(\tau|\lambda_{exc}, \lambda_{inh}^{(1)}, s^{(1)}) \neq f_T(\tau|\lambda_{exc}, \lambda_{inh}^{(2)}, s^{(2)})$ for the same $\lambda_{exc}$ but different inhibitory rates $\lambda_{inh}^{(1)}$

and $\lambda_{inh}^{(2)}$ and/or synchronicities $s^{(1)}$ and $s^{(2)}$. Within this setting, the $\lambda_{exc}$ is communicated through the channel, while $\lambda_{inh}$ and $s$ control the channel characteristics (excitation-driven scenario; see text for more details).

*B. Neuron Model*

We model the postsynaptic neuron as a Hodgkin-Huxley-type (HH-type) neuron with a slow adaptation potassium current, the so-called A-current [37, 38]. This biophysical model has measurable physiological parameters and, unlike the integrate-and-fire models, takes the spike generation biophysics into consideration. The dynamics of the membrane potential, $V(t)$, is governed by

$$C_m \frac{d}{dt} V(t) = -g_L \left( V(t) - E_L \right) - \sum_{int} I_{int}(t) + I_{net}(t), \qquad (1)$$

here $I_{int}(t)$ denotes the neuron's active ionic (the so-called intrinsic) current with Hodgkin-Huxley type kinetics, $I_{net}(t)$ is the synaptic current of the postsynaptic neuron, $g_L$ is the leak conductance ($g_L = 0.05\ mS/cm^2$), $E_L$ is the reversal potential ($E_L = -65\ mV$) of the leak current, $C_m$ is the membrane capacitance ($1\ \mu F/cm^2$), and $t$ is the time. This model considers three types of active ionic currents, $I_{int}(t)$, $int \in \{Na^+, K^+, A\}$: a sodium ($Na^+$) current, $I_{Na^+}$, a delayed-rectifier potassium ($K^+$) current, $I_{K^+}$, and an A-current, $I_A$. Each active ionic current is a voltage-dependent current, which can be described by,

$$I_{int}(t) = g_{int}^{(p)} m_{int}^M(t) n_{int}^N(t) \left( V(t) - E_{int} \right), \qquad (2)$$

where $m_{int}(t)$ and $n_{int}(t)$ are gating variables, and $E_{int}$ and $g_{int}^{(p)}$ are respectively the reversal potential and the peak conductance of active ionic channel. Each active ionic current has specific values of *M* and *N* (see below). For more detailed formulation of these currents as well as the corresponding parameter values of the model, see [37, 38].

*C. Modeling Presynaptic Inputs*

The postsynaptic neuron receives synaptic inputs from *N* presynaptic neurons. We model the synaptic current as

$$I_{net}(t) = \sum_{j=1}^{N} g_j^{(spc)}(t)\left(E_j^{(spc)} - V(t)\right), \tag{3}$$

where $g_j^{(spc)}(t)$ is the synaptic conductance triggered by the spikes of the $j$-th presynaptic neuron, and $E_j^{(spc)}$ is the reversal potential of this synapse ($E_j^{(spc)} = E_{inh} = -80 \; mV$ for inhibitory, $E_j^{(spc)} = E_{exc} = 0 \; mV$ for excitatory). We assume that the synaptic conductance can be described by a linear sum of contributions from all presynaptic spikes. For each spike the $g_j^{(spc)}(t)$ increases instantaneously by weight of $W_j$, and then decays exponentially with a time-constant $\upsilon_j^{(spc)}$. Hence, for $t > 0$, the synaptic conductance is described by

$$\frac{d}{dt}g_j^{(spc)}(t) = -\frac{g_j^{(spc)}(t)}{\upsilon_j^{(spc)}} + W_j R_j(t), \tag{4}$$

where $R_j(t) = \sum_{t_j^s} \delta(t - t_j^s)$ is the instantaneous firing rate of the $j$ th presynaptic neuron, and $t_j^s$ are its spike times. We assume each presynaptic neuron fires a Poisson spike train stochastically and at a fixed rate. In the simulations the synaptic conductance were indeed fluctuating, but at each rate their time-averages (over time $\Delta$), summed over each subpopulation can be computed as

$$\bar{g}_{exc} = N_{exc} \frac{\upsilon_{exc} \lambda_{exc} W_{exc}}{\Delta}, \tag{5-a}$$

$$\bar{g}_{inh} = N_{inh} \frac{\upsilon_{inh} \lambda_{inh} W_{inh}}{\Delta}, \tag{5-b}$$

for the excitatory, $\bar{g}_{exc}$, and inhibitory, $\bar{g}_{inh}$, synapses. $N_{exc}$ and $N_{inh}$ are the number of excitatory and inhibitory neurons. For the derivation of these formulas (and the rest of the paper) we assumed that all the neurons of the same type (excitatory or inhibitory) emit spikes at the same rate $\lambda_j = \lambda_{exc}$ and $\lambda_j = \lambda_{inh}$, and have the synaptic projections with the same synaptic time-constants $\upsilon_j^{(spc)} = \upsilon_{exc}$ and $\upsilon_j^{(spc)} = \upsilon_{inh}$, and weights $W_{exc} = W_j$ and $W_{inh} = W_j$.

*D. Modeling Neuronal Synchronization*

Synchronous spiking of neurons is a ubiquitous phenomenon in neuronal networks. This phenomenon has been observed in different brain regions, during both spontaneous activity, and evoked activity in the presence of stimulus or when performing a task [39-46]. Several biophysical mechanisms, underlying this phenomenon, have been already identified. These include synchronization via i) gap junctions [47], ii) GABAergic interneurons [48], iii) ephaptic coupling [49-52], iv) common input [44, 46, 53], and v) a self-organized process [54]. Moreover, the degree of the synchronization (i.e. the portion of neurons participating in the synchronous event) can be modulated by e.g., the features of the external stimulus (e.g. orientation, intensity level, or context), switching attention [40, 45, 55-57], or the internal state of the network [58].

Nonetheless, to model the synaptic activity with a controlled synchronization level we use a simple procedure for generating spikes and synchronous patterns. In other words, we forgo modeling such afore-mentioned, detailed mechanisms of the neuronal synchronization. This is because to establish the information theoretic framework presented in this work, we only needed to use the synchronous firing patterns which can provide a good approximation to the output of such mechanisms. To this end, we model the occurrence of synchronous spiking events as another Poisson process, i.e., whenever this process "fires a spike" it shall be a population spike with many (or potentially all) neurons in the population participating. Our goal is to control synchronization level in each presynaptic subpopulation (excitatory or inhibitory) independent from its mean firing rate. As a consequence, in order to keep the mean rate constant between the spiking activities in i) the absence ('old') and ii) the presence ('new') of the synchronous events, the spiking probability of the neurons needs to be lowered when synchronous events are injected. Note that this reduction is applied to all neurons of each subpopulation, separately. Overall, the new lowered firing rate in the presence of synchronous events for each presynaptic neuron can be computed as $\lambda_j^{(\text{new})} = \lambda_j^{(\text{old})} - s\lambda_j^{(\text{sync})}$, where $\lambda_j^{(\text{old})}$ is the rate in the absence of synchronous events, $\lambda_j^{(\text{sync})}$ is the rate of synchronous events, and $0 < s < 1$ denotes the fraction of randomly chosen presynaptic spike trains participating in the synchronous events (e.g., $sN_{exc}$ excitatory neurons are participating), and thus determines the synchronization level in the corresponding subpopulation. This way of modeling the neuronal synchronization is similar to the previously established approaches, e.g. see [53, 59]. Moreover, note that this model enabled us to simulate the synchronization events with specific synchronization levels, instead of having variability in the synchronization levels during simulation time. In turn, in our results, this modeling approach enabled us to calculate the relationship between the synchronization level and the mutual information per unit cost more accurately (see Results). In this paper, we consider the synchronization only in the presynaptic excitatory subpopulation.

*E. Modeling Balanced Regimes*

Balanced regimes of excitatory and inhibitory synaptic inputs [10, 60] are thought to play an important role in information processing of cortical neurons *in vivo* [10, 61, 62]. For instance, recently it has been reported that these regimes can potentially promote both coding efficiency and energy efficiency of neural computation [63]. Under such a regime, the co-activation of excitation and inhibition can hover the membrane potential of neurons somewhat below the firing threshold [60, 64], e.g. as in the so-called Up-state of subthreshold potential [39, 65], where they will fire in response to small, but sufficiently large, depolarizing fluctuations in their synaptic inputs [10]. Accordingly, these neurons exhibit highly irregular firing patterns which are statistically similar to Poisson spike trains [60, 66]. To implement these experimentally reported regimes in our modeling framework of the synaptic inputs, we develop two simulation scenarios called excitation-driven and fluctuation-driven scenarios as follows (the full list of parameter values used in these scenarios can be found in Table 1).

**Excitation-driven simulation scenario.** In this scenario, we only change the firing rate of the excitatory neurons and keep the firing rate of the inhibitory neurons constant. Then, for each synchronization level, we obtain the firing rate and ISI statistics of postsynaptic neuron as a function of the mean presynaptic excitatory firing rate $\lambda_{exc}$, via simulations. Of note, different levels of $\lambda_{exc}$ can be due to different baseline (spontaneous) firing activity level of the neurons, or to be achieved e.g. in a task- or stimulus-dependent manner (e.g. by using different stimulus contexts, intensities, or frequencies); see for example [67-70]. We formulate this scenario as follows. First, we define a constant input current $I(\infty) = -g_L(V(\infty) - E_L)$, which in the absence of active ionic currents (see (1)) can produce an asymptotic voltage of the firing threshold of the full HH-type neuron model; i.e. $V(\infty) = -48\,mV$. We set this current equal to the summation of the time-averaged presynaptic currents of all excitatory ($\bar{I}_{exc}$) and inhibitory ($\bar{I}_{inh}$) synapses, i.e. $I(\infty) = \bar{I}_{exc} + \bar{I}_{inh}$. We then find the desired parameter values of the corresponding synaptic input currents. Note that within our derivation, we also make two biologically plausible assumptions: (i) the total mean firing rate summed over all the presynaptic excitatory neurons, $\lambda_{exc}^{(tot)} = N_{exc}\lambda_{exc}$, is equal to that of the presynaptic inhibitory neurons, $\lambda_{inh}^{(tot)} = N_{inh}\lambda_{inh}$, and (ii) $\bar{I}_{exc} = 2I(\infty)$, i.e., without inhibition the excitatory drive would push the membrane potential way above the firing threshold. Overall, this formalism results in the constant synaptic conductance values per synapse (i.e. the weights $W_{exc}$ and $W_{inh}$), which are independent of the synchronization level:

$$W_{exc} = 1000 \frac{\bar{I}_{exc}}{\upsilon_{exc} \lambda_{exc}^{(tot)} \left(V(\infty) - E_{exc}\right)}, \tag{6-a}$$

$$W_{inh} = 1000 \frac{\bar{I}_{inh}}{\upsilon_{inh} \lambda_{inh}^{(tot)} \left(V(\infty) - E_{inh}\right)}, \tag{6-b}$$

where by setting $\lambda_{exc}^{(tot)} = \lambda_{inh}^{(tot)} = 5000\,\text{Hz}$ (consistent with [71]) for $N_{exc} = 160$ ($N_{exc} = 80\%$ of $N_{tot}$, where $N_{tot} = 200$) and $N_{inh} = 40$ (i.e. $N_{inh} = 20\%$ of $N_{tot}$) synapses (consistent with [72]) with $\upsilon_{exc} = \upsilon_{inh} = 10\,m\sec$ (consistent with [73]), we obtained the weights $W_{exc}$ and $W_{inh}$ as $7.0833 \times 10^{-4}$ and $5.3125 \times 10^{-4}\,mS/cm^2$. Then, we fixed the firing rate of the presynaptic inhibitory neurons ($\lambda_{inh} = \lambda_{inh}^{(tot)} / N_{inh} = 125\,\text{Hz}$) and simulated the full HH-type model for different rates of the presynaptic excitatory neurons, $\lambda_{exc}$, as well as different synchronization levels, $s$. No additional background inputs or sources of noise were modeled or simulated.

**Fluctuations-driven Simulation Scenario.** This scenario corresponds to a setting, where the presynaptic inhibitory rates co-vary with presynaptic excitatory rates such that they approximately cancel out their postsynaptic effects, and the postsynaptic response is mainly due to an increase in the fluctuations caused by increasing the excitatory and inhibitory inputs together. More specifically, we enforce the currents $\bar{I}_{exc}$ and $\bar{I}_{inh}$ (see above) to cancel each other:

$$\bar{I}_{exc} = -\bar{I}_{inh}, \tag{7-a}$$

$$\bar{g}_{exc}\left(V(\infty) - E_{exc}\right) = -\bar{g}_{inh}\left(V(\infty) - E_{inh}\right), \tag{7-b}$$

$$\lambda_{exc} \frac{N_{exc} W_{exc} \upsilon_{exc}}{\Delta}\left(V(\infty) - E_{exc}\right) = -\lambda_{inh} \frac{N_{inh} W_{inh} \upsilon_{inh}}{\Delta}\left(V(\infty) - E_{inh}\right), \tag{7-c}$$

$$\lambda_{inh} = -\lambda_{exc} \frac{N_{exc} W_{exc} \upsilon_{exc}}{N_{inh} W_{inh} \upsilon_{inh}} \frac{\left(V(\infty) - E_{exc}\right)}{\left(V(\infty) - E_{inh}\right)}, \tag{7-d}$$

where $W_{exc} = 7.0833 \times 10^{-4}\,mS/cm^2$ (see the excitation-driven scenario, above), $W_{inh} = \gamma W_{exc}$ with now $\gamma = 5$, $N_{exc} = 320$ (i.e. $N_{exc} = 80\%$ of $N_{tot}$, where $N_{tot} = 400$) and $N_{inh} = 80$ (i.e. $N_{inh} = 20\%$ of $N_{tot}$) (consistent with [72]), and as above $\upsilon_{exc} = \upsilon_{inh} = 10\ m\sec$ and $V(\infty) = -48\ mV$ (according to the neuron model). The last equation gives the firing rate of presynaptic inhibitory neurons as a linear function of the excitatory rate.

Table 1. Parameter values used in the two simulation scenarios of synaptic inputs.

| Simulation Scenario | Parameters | Units |
|---|---|---|
| Excitation-driven | $\lambda_{exc}^{(tot)} = 5000$ | Hz |
| | $\lambda_{inh}^{(tot)} = 5000$ | Hz |
| | $W_{exc} = 7.0833 \times 10^{-4}$ | mS/cm$^2$ |
| | $W_{inh} = 5.3125 \times 10^{-4}$ | mS/cm$^2$ |
| | $N_{tot} = 200$ | - |
| | $N_{exc} = 160$ | - |
| | $N_{inh} = 40$ | - |
| | $\lambda_{inh} = 125$ | Hz |
| | $V(\infty) = -48$ | mV |
| Fluctuation-driven | $W_{exc} = 7.0833 \times 10^{-4}$ | mS/cm$^2$ |
| | $N_{tot} = 400$ | - |
| | $N_{exc} = 320$ | - |
| | $N_{inh} = 80$ | - |
| Both scenarios | $V(\infty) = -48$ | mV |
| | $\upsilon_{exc} = 10$ | ms |
| | $\upsilon_{inh} = 10$ | ms |
| | $E_L = -65$ | mV |
| | $g_L = 0.05$ | mS/cm$^2$ |
| | $C_m = 1$ | µF/cm$^2$ |
| | $E_{exc} = 0$ | mV |
| | $E_{inh} = -80$ | mV |

### F. Parametric Distributions

To enable (semi) analytic analyses of neuronal communications in this framework, we fit the postsynaptic ISI distribution, computed from our computer simulations, with parametric distributions. To do so, we first fitted the postsynaptic ISI distribution with *multiple* distributions to obtain a closed form probability distribution function (PDF) for postsynaptic ISI distribution with given values of input rate and controlling parameters, $f_T(\tau|\lambda_{exc},\lambda_{inh},s)$. For each combination of input rates, $\lambda_{exc}$, and controlling parameters, $\lambda_{inh}$ and $s$, we derive a separate fit. To gain insights into how the changes of input rates and synchronization affect the postsynaptic ISI distribution, it is helpful to have an interpretation available of how the parameters of the distributions affect its shape: The distributions we selected are parameterized by *scale*, *shape*, and *location* parameters (see Table S1 in supplementary materials). The larger the value of the *scale parameter*, the more the distribution is spread out. A change to a location parameter (a scalar or a vector) simply shifts the distribution. Finally, a *shape parameter* is neither stretching nor shifting a distribution but affects only its shape, for example, affecting how skewed a distribution is. In the next section we use parametric distributions to fit them to the conditional distribution of postsynaptic ISI for given value of presynaptic excitatory firing rate.

### G. *Statement of Optimization Problem*

The synchronization level (the control parameter) affects the conditional distribution of postsynaptic ISI given presynaptic firing rate, which in turn affects the transmitted information per unit cost. Here we *define* an optimization problem to determine the "optimal synchronization level" within the modeled system, in terms of energy efficiency. For an optimal synchronization level, the following equation needs to be satisfied:

$$s^* = \arg\max \mathcal{I}_{bpj}(s), \tag{8}$$

where $\mathcal{I}_{bpj}(s)$ is average mutual information per unit energy expenditure and is given by

$$\mathcal{I}_{bpj}(s) = \max_{F_{\Lambda_{exc}|S}(\lambda_{exc}|S=s)} \frac{\mathcal{I}}{E(e(\tau)|S=s)},$$
$$\text{s.t.} \tag{9}$$
$$F_{\Lambda_{exc}|S}(\lambda_{exc}|S=s) = \Pr(\Lambda_{exc} < \lambda_{exc}|S=s),$$

where $\mathcal{I}$ is the average mutual information (see following text), $E(\cdot)$ denotes expectation operation and $e(t)$ denotes the energy expenditure of a neuron *during* the postsynaptic ISI of duration $T$, and is given by [24, 34]

$$e(\tau) = C_0 + C_1\tau, \tag{10}$$

where, $C_0$ and $C_1$ are constants [24], and $F_{\Lambda_{exc}|S}(\lambda_{exc}|s)$ denotes the cumulative distribution function (CDF) of $\lambda_{exc}$ for a given value of $S = s$ and indicates the distribution of channel's input . The average mutual information is given by

$$\mathcal{I} = \frac{1}{L} \lim_{N \to \infty} \mathcal{I}\left(\Lambda_{exc,1},...,\Lambda_{exc,L};T_1,...,T_L \big| S = s\right), \tag{11}$$

where $\Lambda_{exc,i}$ and $T_i, i \in \{1,...,L\}$ denote the firing rate of presynaptic neurons and ISI of postsynaptic neuron, and $L$ is the number of postsynaptic ISIs. To solve the optimization problem (8), we first need to solve the optimization problem (9) by considering $S$ as a parameter.

## III. RESULTS

### A. Conditional Distribution of Postsynaptic ISI

We first simulated the one-compartment HH-type neuron model driven by presynaptic excitatory and inhibitory inputs for different values of the input rate $\lambda_{exc}$ of the excitatory neurons. The rate $\lambda_{inh}$ of the inhibitory neurons was kept constant (excitation-driven scenario). The mean firing rates of postsynaptic neuron are shown in Figure 2A. The input-output functions are approximately linear due to the adaptation current included in the neuron model [74]. The mean firing rates of postsynaptic neuron are very similar for different synchronization levels in the presynaptic excitatory population, except for the range of input rates close to the onset-threshold of the mean output rate (Figure 2A, inset). For the same input rates $\lambda_{exc}$ of approximately 15-20 Hz, the neuron responds with higher output rates for higher synchronization levels, which can be explained by the higher levels of voltage fluctuations experienced by the postsynaptic neuron: When a larger fraction of presynaptic excitatory neurons synchronize their discharges, the fluctuations in the postsynaptic membrane potential are larger and therefore can occasionally trigger spikes. Although the mean firing rates do not differ much for different synchronization levels, the variability of the postsynaptic firing rate increases with synchronization, which is shown in Figure 2B in terms of the coefficient of variation of the simulated postsynaptic ISIs(CV = $std(T)/mean(T)$). Higher CVs corresponds to more irregular firing (compare CVs for 90% and 70% synchronization levels in Figure 2B). Moreover, the dependence of the CV on the input rate $\lambda_{exc}$ is concave with a maximum at about 20 Hz. Figure 3C shows the conditional entropy of postsynaptic ISI for given values of firing rate of presynaptic neurons, $h\left(T \mid \Lambda_{exc} = \lambda_{exc}, \Lambda_{inh} = \lambda_{inh}\right)$, in terms of synchronization level for different levels of $\lambda_{exc}$ with $\lambda_{inh} = 125$ Hz. There, it can be seen that $h\left(T \mid \Lambda_{exc} = \lambda_{exc}, \Lambda_{inh} = \lambda_{inh}\right)$ increases for higher values of $\lambda_{exc}$. Moreover, it is obvious (Figure 2C) that $h\left(T \mid \Lambda_{exc} = \lambda_{exc}, \Lambda_{inh} = \lambda_{inh}\right)$ has a minimum in terms of the synchronization level. Further characterizations of the postsynaptic ISI distributions for different synchronization levels and input rates are shown in Figures 2D-F in terms of the CV (Figure 2D), the skewness (skewness= $E\left(T - mean(T)\right)^3 / \left(std(T)\right)^3$, Figure 2E), and the kurtosis (kurtosis= $E\left(T - mean(T)\right)^4 / \left(std(T)\right)^4$, Figure 2F) of the postsynaptic ISI distribution. As it can be observed the skewness and kurtosis of postsynaptic ISI are respectively 0 and 3 for synchronization levels less than 30%, hence, postsynaptic ISIs are normally distributed for synchronization levels less than 30%. It is worth noting in figures 2B-F that the full postsynaptic ISI distribution is strongly affected by synchronization level. Therefore, these results (Figure 2) imply that synchronization is indeed an effective control parameter for postsynaptic ISI.

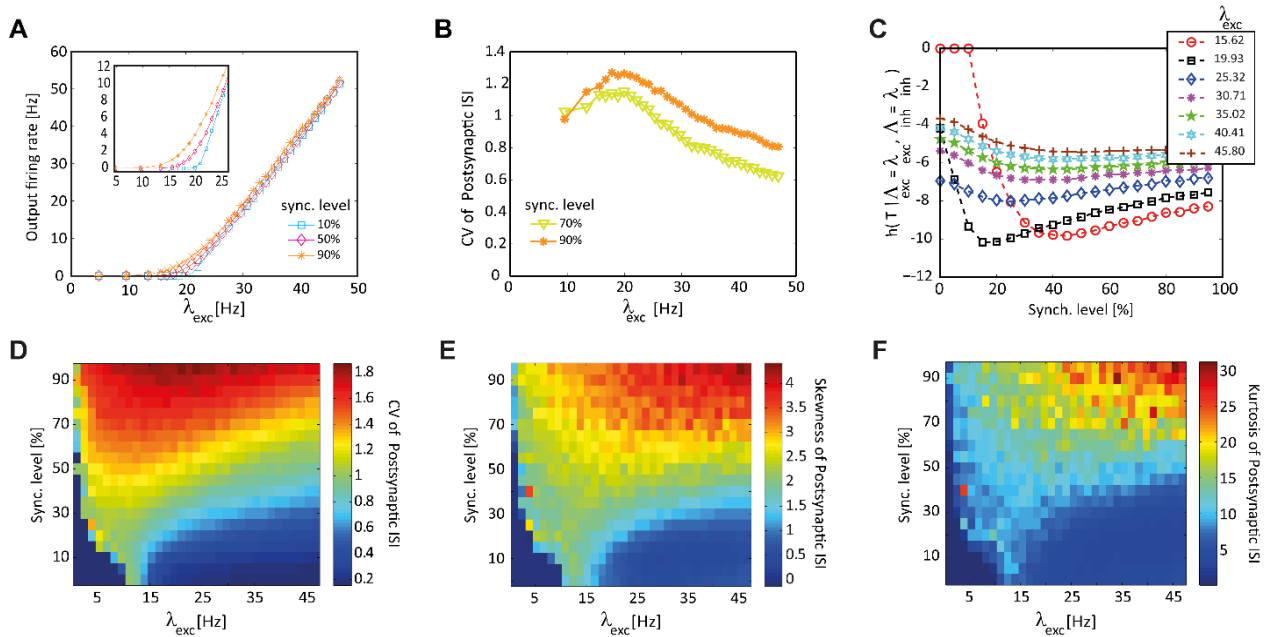

**Figure 2.** Input-output transformation of the simulated neuron model in the excitation-driven scenario: *A,* Mean of the output (i.e. postsynaptic) firing rate as a function of the input (i.e. presynaptic) firing rate $\lambda_{exc}$, for different synchronization levels. The rate of the inhibitory neurons was kept constant (excitation-driven scenario, see Methods). The inset displays the zoom into these output rates, showing that for higher synchronization levels (and hence higher postsynaptic voltage fluctuation levels) the onset-threshold of the mean output rate is lowered. *B,* The regularity of firing is quantified with the coefficient of variation (CV), which decreases for higher input rates (firing becomes more regular). Firing is more irregular for high synchronization level. *C,* Conditional entropy of the postsynaptic ISI for given values of firing rates of presynaptic neurons, $h\left(T|\Lambda_{exc}=\lambda_{exc},\Lambda_{inh}=\lambda_{inh},S=s\right)$, in terms of presynaptic excitatory firing rate ($\lambda_{exc}$) with $\lambda_{inh}=125$ Hz for different synchronization levels, $s$. *D-F,* More fine-grained characterization of the postsynaptic ISI distribution for different input rates and synchronization levels demonstrated by *D,* the output CV, *E,* the output skewness, and *F,* the output kurtosis.

We now set up the fluctuations-driven scenario to check whether our main results (Figure 2) are dependent on the transmitting regime of the presynaptic inputs. Note that in the fluctuations-driven scenario $\lambda_{exc}$ and $\lambda_{inh}$ are varied together (see Methods). Figure 3A shows the mean of output firing rate in Hz in terms of $\lambda_{exc}$ for different synchronization levels. It can be observed that the mean output firing rate increases by increasing $\lambda_{exc}$, but not by changing the synchronization level. Moreover, it can be observed that by increasing the synchronization level, the onset-threshold of the mean output rate is reduced (Figure 3A, inset). Figure 3B shows CV of postsynaptic ISIs in terms of $\lambda_{exc}$ for different synchronization levels. We found that increasing the synchronization level, increases the CV of postsynaptic ISIs, which implies more dispersion in postsynaptic ISI distribution. Figure 3C shows the conditional entropy of postsynaptic ISI for given values of firing rate of presynaptic neurons, $h\left(T|\Lambda_{exc}=\lambda_{exc},\Lambda_{inh}=\lambda_{inh}\right)$, in terms of synchronization level for different levels of $\lambda_{exc}$ and $\lambda_{inh}$. It is evident that $h\left(T|\Lambda_{exc}=\lambda_{exc},\Lambda_{inh}=\lambda_{inh}\right)$ increases for higher values of $\lambda_{exc}$. It is also clear in Figure 3C that $h\left(T|\Lambda_{exc}=\lambda_{exc},\Lambda_{inh}=\lambda_{inh}\right)$ has a minimum in terms of the synchronization level. Figures 3D-E show the CV, skewness and kurtosis of postsynaptic ISI in terms of $\lambda_{exc}$ and synchronization level, where it can be seen that increasing the synchronization level leads to an increase in all these statistical parameters. Moreover, Figures 3E and 3F show that for synchronization levels less than 30%, the kurtosis and skewness of postsynaptic ISI are about 3 and 0, respectively.

Collectively, these results (Figures 2 and 3) indicate that changing the synchronization level, does not affect the mean of firing rate in postsynaptic neuron; however, it does affect the overall distribution as it influences the higher order moment of ISI. Moreover, the conditional entropy of postsynaptic ISI for given values of $\lambda_{exc}$ and $\lambda_{inh}$ in terms of synchronization level is a convex function of synchronization level. This in principle indicates that the average mutual information should have a maximum in terms of the synchronization level.

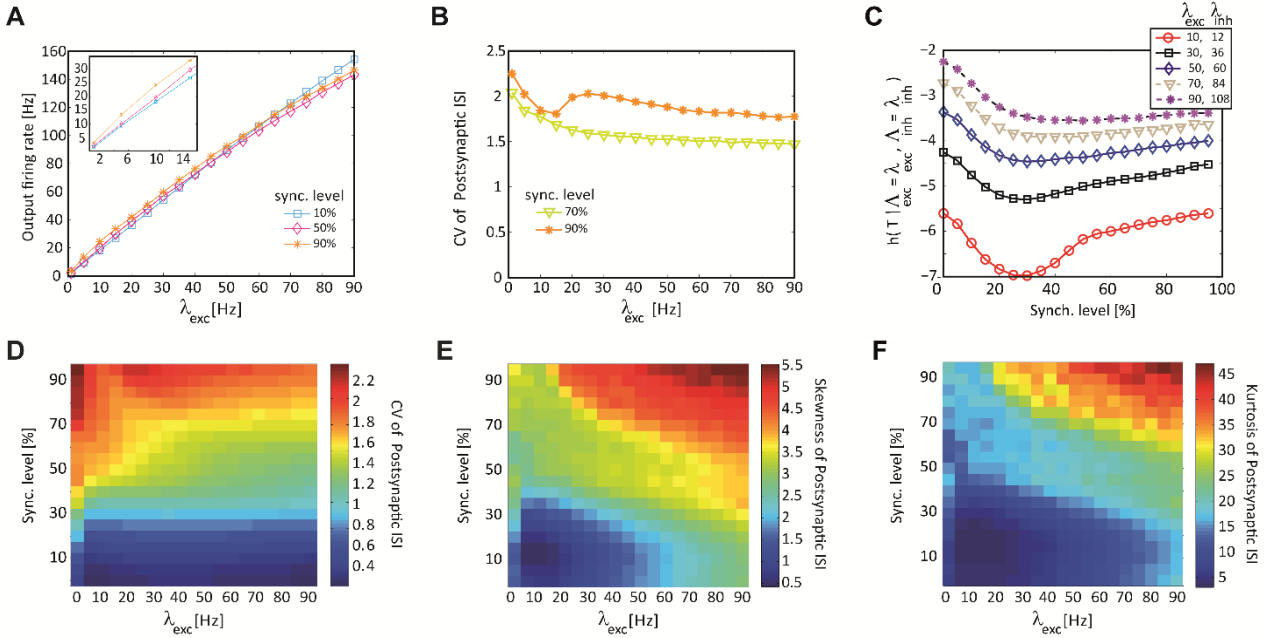

**Figure 3.** Input-output transformation of the simulated neuron model in fluctuation-driven scenario: The same format is used as in Figure 2. *A,* Mean of output (i.e. postsynaptic) firing rate in terms of input (i.e. presynaptic) firing rate $\lambda_{exc}$ for different synchronization levels, *B,* CV of postsynaptic ISI in terms of $\lambda_{exc}$ for different synchronization levels. *C,* Conditional entropy, $h\left(T \mid \Lambda_{exc} = \lambda_{exc}, \Lambda_{inh} = \lambda_{inh}\right)$, in terms of synchronization level for different values of $\lambda_{exc}$ and $\lambda_{inh}$. *D-F,* The statistical parameters of postsynaptic ISI in terms of $\lambda_{exc}$ for different levels of synchronization: *D,* CV, *E,* Skewness, *F,* Kurtosis.

The objective function in (9) corresponds to the average mutual information per unit energy expenditure, and we show in the sequel that it can be simplified to conditional entropy of ISI with two constraints. Hence, we here study the conditional entropy of ISI for given values of presynaptic inhibitory and excitatory firing rates. Comparing the results of the excitation-driven and fluctuations-driven scenarios (Figures 2C and 3C), we conclude that the conditional entropy of ISI has a global minimum at certain synchronization levels in both transmitting regimes of synaptic inputs. The minimum value of the conditional entropy corresponds to the minimum uncertainty about post synaptic ISI, when $\lambda_{exc}$ and $\lambda_{inh}$ are known. Therefore, we focus only on the excitation-driven scenario in the rest of the paper, as a representative simulation scenario. To further facilitate the analysis, in the following, we describe the postsynaptic ISI distribution in terms of parameterized distributions.

We use Kolmogorov-Smirnov test ($\alpha = 5\%$) to identify the proper parametric distribution to fit the postsynaptic ISIs obtained from the simulations. The null hypothesis is that the postsynaptic ISIs were generated from the continuous distribution that we test. The alternative hypothesis is that the postsynaptic ISIs were generated from a different continuous distribution. For the set of rates $\lambda_{exc} \in \{0, 4, ..., 48\}$ Hz, let $q(i)$, $i \in \{1, 2, ..., H_{exc}\}$, where $H_{exc} = 13$, be an indicator variable that shows whether the null hypothesis is rejected, $q(i) = 1$, otherwise $q(i) = 0$. Let

$$RKS_{test} = \left(1 - \frac{\sum_{i=1}^{N_{exc}} q(i)}{H_{exc}}\right) \times 100 \qquad (12)$$

be the ratio of Kolmogorov Smirnov tests, where the null hypothesis could *not* be rejected. Figure 4A shows $RKS_{test}$ for the six different distributions that we tested (See Table S1 in supplementary materials). We found that, while for synchronization levels below 40%, the Gamma, Log-Normal, Weibull, and Generalized Extreme Value (GEV) distributions are all proper fits to the postsynaptic ISI histograms (for different values of the input rate $\lambda_{exc}$), but for higher synchronization level, only the GEV distribution is able to describe the simulated postsynaptic ISI histograms well. To better understand these differences, we investigated the postsynaptic ISI histograms with their approximations to the GEV and the Gamma distributions, which provided better fits. Figures 4B, and 4C show the postsynaptic ISI histograms and the fitted GEV distribution for a fixed $\lambda_{exc}$ of 36 Hz as an exemplified input rate, and different synchronization levels (Figure 4B for $s \leq 40\%$, Figure 4C for $s > 40\%$). The Gamma distribution fits the simulations well for synchronization levels below 40% (Figure 4D), but not at higher synchronization levels.

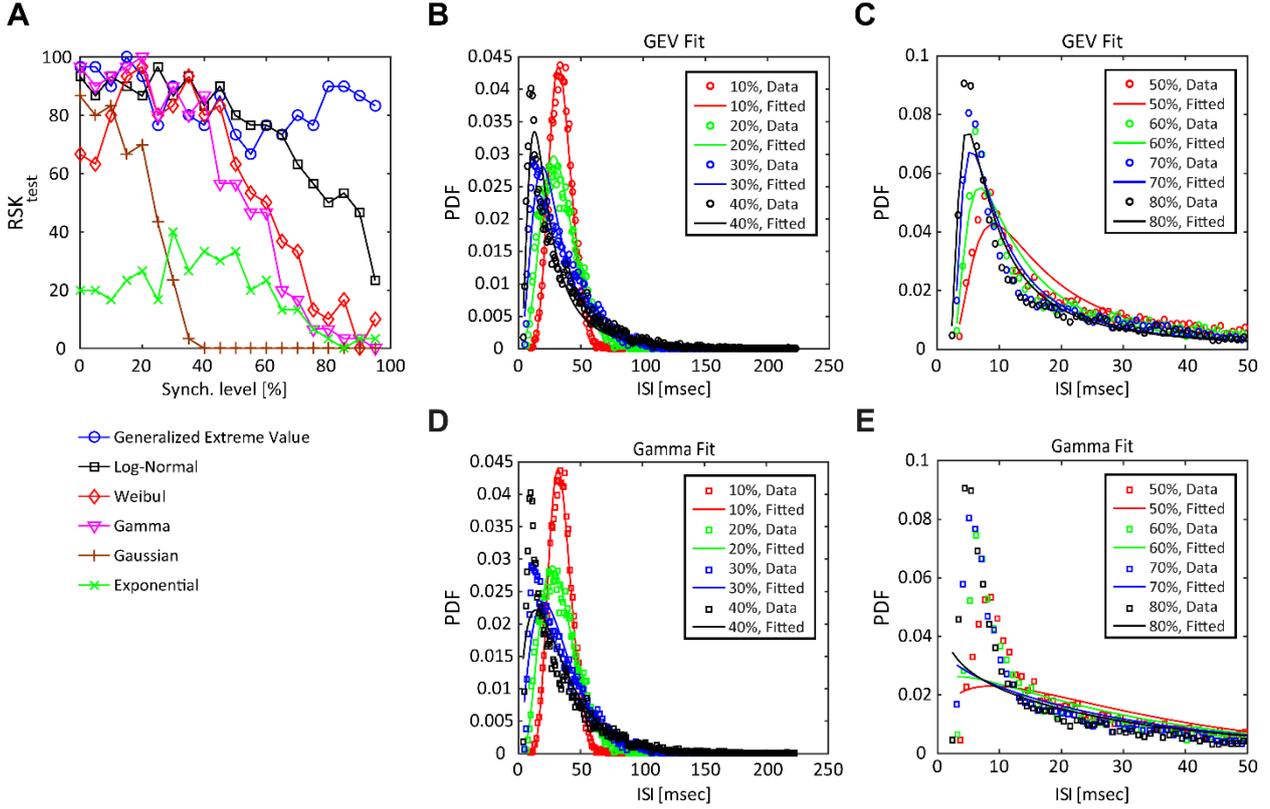

**Figure 4.** Results of fitting the simulation results with parametric distributions for multiple input rates and synchronization levels: *A,* For each synchronization level we tested if the postsynaptic ISI histogram for different input rates, $\lambda_{exc}$, can be described with one of the listed parametric distributions (see legend). $RKS_{test}$ is the ratio of Kolmogorov–Smirnov test (with $\alpha = 0.05$), which determines to what extent the ISI histogram is well described by the corresponding parametric distribution. The exponential distribution (green line) is in general a poor fit to the simulated postsynaptic ISIs. Note that only the generalized extreme value (GEV) distribution (blue line) describes the data well for all synchronization levels. *B, C,* Postsynaptic ISI distributions for a fixed input rate $\lambda_{exc} = 36$ Hz and different synchronization levels: *B,* $0\% \leq s \leq 40\%$, and *C,* $50\% \leq s \leq 80\%$. Shown are the simulated results and the GEV distribution fit (circles). *D and E:* same as *B* and *C,* but for the Gamma distribution. The data were generated under a simulation time of 2000 sec and synchronization levels of 0%, 5%, ..., 95%.

To this end, we found the GEV distribution to be a good candidate for modeling the conditional distribution of postsynaptic ISI. We then investigated how the parameters of the GEV distribution depend on the input rate and synchronization level. More specifically, we aimed at expressing these dependencies using linear or quadratic functions, which we will use in the subsequent (semi)analytical calculations. For a given value of $\lambda_{exc}$, the PDF of the GEV distribution for the location parameter $k^{(\text{gev})} \neq 0$ is given by

$$f_{T|\Lambda_{exc},S}(\tau|\lambda_{exc},s) = \mathbb{GEV}\left(\mu^{(\text{gev})}(s,\lambda_{exc}), \sigma^{(\text{gev})}(s,\lambda_{exc}), k^{(\text{gev})}(s,\lambda_{exc})\right) \tag{13}$$

where the location, shape and scale parameters, $k^{(\text{gev})}(s, \lambda_{exc})$, $\sigma^{(\text{gev})}(s, \lambda_{exc})$ and $\mu^{(\text{gev})}(s, \lambda_{exc})$ are fitted by quadratic functions (note that linear functions are *not* sufficient to describe the parameters' dependence on the input rate)

$$k^{(\text{gev})}(s, \lambda_{exc}) = \sum_{i=0}^{2} k_i^{(\text{gev})}(s) \lambda_{exc}^i \tag{14-a}$$

$$\sigma^{(\text{gev})}(s, \lambda_{exc}) = \sum_{i=0}^{2} \sigma_i^{(\text{gev})}(s) \lambda_{exc}^i \tag{14-b}$$

$$\mu^{(\text{gev})}(s, \lambda_{exc}) = \sum_{i=0}^{2} \mu_i^{(\text{gev})}(s) \lambda_{exc}^i \tag{14-c}$$

where $k_i^{(\text{gev})}(s)$, $\sigma_i^{(\text{gev})}(s)$ and $\mu_i^{(\text{gev})}(s)$, $i \in \{0,1,2\}$ are synchrony-dependent coefficients of the fitted quadratic functions. Figures 5A-C show the dependence of $k^{(\text{gev})}(s, \lambda_{exc})$, $\sigma^{(\text{gev})}(s, \lambda_{exc})$ and $\mu^{(\text{gev})}(s, \lambda_{exc})$ on $\lambda_{exc}$. Unfortunately, based on this formulation, due to some technical limitations, we cannot find closed form expression for the solution of the optimization problem in (9) with the GEV distribution. Accordingly, hereafter, we repeated the same procedure with the Gamma distribution, which, as we already showed (Figure 4), is a good fit to the simulated data for synchronization levels less than 40% only. In other words, in the following we investigate how the parameters of the Gamma distribution depend on the input rate and synchronization level.

The PDF of the Gamma distribution of postsynaptic ISI is given by

$$f_{T|\Lambda_{exc},S}(\tau | \lambda_{exc}, s) = \mathbb{GAM}\left(b^{(\text{gam})}(s, \lambda_{exc}), m^{(\text{gam})}(s, \lambda_{exc})\right) \tag{15}$$

where, the inverse scale and shape parameters, $b^{(\text{gam})}(s, \lambda_{exc})$ and $m^{(\text{gam})}(s, \lambda_{exc})$ are fitted with a quadratic and a linear function, respectively, as follows

$$b^{(\text{gam})}(s, \lambda_{exc}) = \sum_{i=0}^{1} b_i^{(\text{gam})}(s) \lambda_{exc}^i, \tag{16-a}$$

$$m^{(\text{gam})}(s, \lambda_{exc}) = \sum_{i=0}^{2} m_i^{(\text{gam})}(s) \lambda_{exc}^i, \tag{16-b}$$

where $b_i^{(\text{gam})}(s)$ and $m_i^{(\text{gam})}(s)$ are synchrony-dependent fitting coefficients. Figure 6 shows the dependencies of the parameters of the Gamma distribution on the input rate. Figures 6A and 6B show the shape and scale parameters ($m^{(gam)}(s,\lambda_{exc})$ and $b^{(gam)}(s,\lambda_{exc})$), respectively. We found that $m^{(gam)}(s,\lambda_{exc})$ is well fitted by a quadratic function (See Figure 6A). To fit $b^{(gam)}(s,\lambda_{exc})$ using a linear function is sufficient (See Figure 6B). Using these fitted polynomials, we are ready for solving optimization problem in (9) in the next subsection.

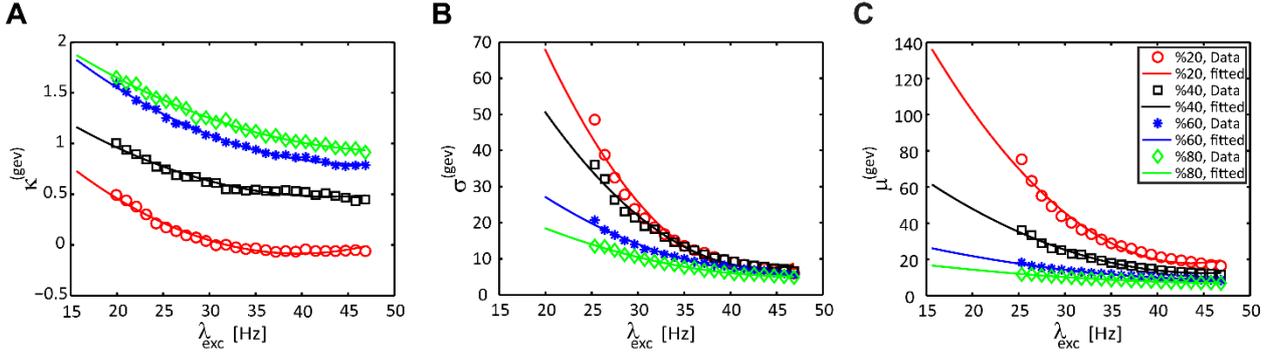

**Figure 5.** Parameters of the GEV distribution fit to the postsynaptic ISIs in terms of the input rates, $\lambda_{exc}$. *A-C*, the estimated parameter values by maximum likelihood estimators are shown with markers and the quadratic fits to the estimated values are shown by solid lines. *A,* Estimated location parameter, $k^{(gev)}(s,\lambda_{exc})$, *B*, estimated shape parameter, $\sigma^{(gev)}(s,\lambda_{exc})$, and *C*, estimated scale parameter, $\mu^{(gev)}(s,\lambda_{exc})$, of the fitted GEV distribution, for different synchronization levels (20-80%).

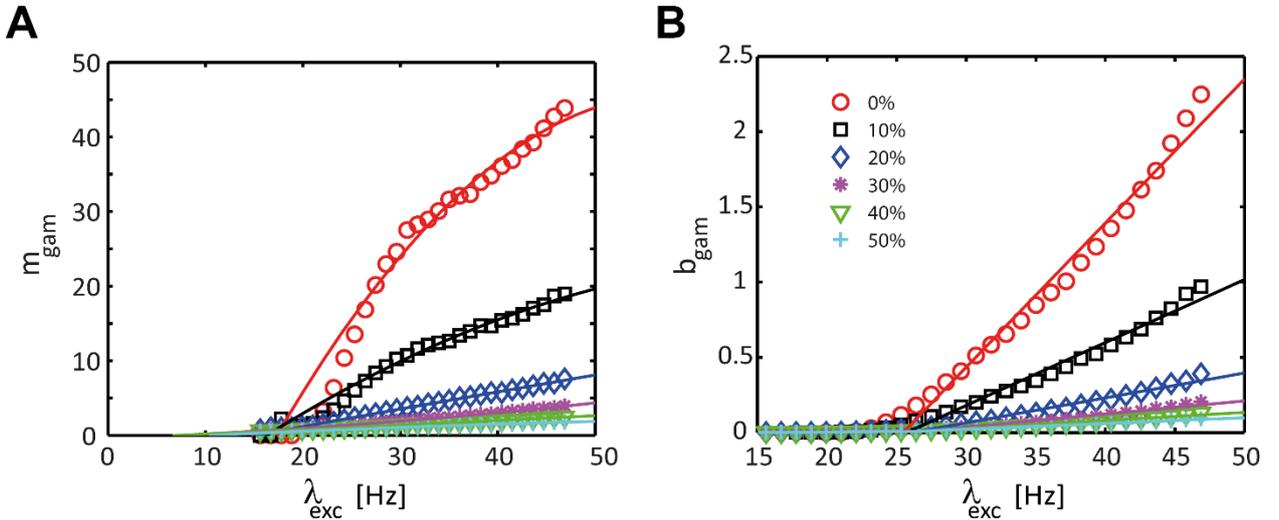

**Figure 6.** Parameters of the Gamma distribution fit to the postsynaptic ISI in terms of input rates, $\lambda_{exc}$. *A*, *B*, the estimated parameter values by maximum likelihood estimators are shown with markers and the polynomial fits to the estimated values are shown by solid lines. *A,* estimated shape parameter of the fitted Gamma distribution, $m^{(gam)}(s,\lambda_{exc})$, and its quadratic fit. *B,* estimated scale parameter of the fitted Gamma distribution, $b^{(gam)}(s,\lambda_{exc})$, and its linear fit, for different synchronization levels (0-50%).

### B. *Optimal Input Rate Distributions*

In this Section, we find a closed form expression for the distribution of presynaptic firing rate, which maximizes the average mutual information per energy expenditure unit ($\mathcal{I}_{bpj}(s)$ in (9)), for given synchronization level. We determine an equivalent problem for solving the optimization problem in (9), which is relatively easier to solve than the original problem. For technical reasons finding a closed form expression for the optimization problem (9), with $f_{T|\Lambda_{exc},S}(\tau|\lambda_{exc},s)$ fitted with Gamma distribution as in (15), is not possible in general. However, as we shall elaborate below, this can be still possible for synchronization levels between 15% and 50%. Principally, in this case the problem reduces to finding the CDF of postsynaptic ISIs, $F_{T|S}(\tau|s)$, which maximizes the entropy of postsynaptic ISIs, $h(T|s)$, subject to a set of constraints (see the following equation). Upon obtaining $F_{T|S}(\tau|s)$ for feasible values of the constraints in (9), we then seek the corresponding optimized $F_{\Lambda_{exc}|S}(\lambda_{exc}|s)$.

Specifically, as reported in Appendix A of the Supplementary Material, the optimization problem in (9), can be simplified as follows

$$\max_{F_{T|S}(\tau|s)} h(T|s), \tag{17-a}$$

s.t.
$$F_{T|S}(\tau|s) = \Pr(T < \tau|s), \tag{17-b}$$
$$E(T|s) = g_0, \tag{17-c}$$
$$E(\log_e T|s) = g_1, \tag{17-d}$$

where $g_0$ denotes the constraint on $E(T|s)$, which comes from the expression for the energy in the form of $e(\tau) = C_0 + C_1 \tau$. In (17-c), $g_0$ denotes the constraint on $E(\log_e T|s)$ stemming from the expression for the mutual information between two successive postsynaptic ISI and can be approximated by [24]

$$\mathcal{I}(T_1; T_2) \approx E\{-\kappa \log(T_1) + C\}, \text{ for small } T_1 \tag{18}$$

where, $C$ and $\kappa$ are constant values dependent on the constant normalized covariance of postsynaptic ISI, $T$, and $T_1$ and $T_2$ are two successive postsynaptic ISIs. The steps to prove that the optimization problem in (17-a) is equivalent to (9) includes computing the Lagrangian function of (11) using the fitting parameters of (16-a) and changing the variable $\tau$ to $x = \left(b_1^{(\text{gam})}(s)\lambda_{exc} + b_0^{(\text{gam})}(s)\right)\tau$ (details are available in supplementary material, Appendix A). The optimized distribution for $f_{T|S}(\tau|s)$ is a Gamma distribution [75] as

$$f_{T|S}(\tau|s) = \frac{\beta^{\kappa}\tau^{\kappa-1}e^{-\beta\tau}}{\Gamma(\kappa)}u(t), \tag{19}$$

where $\beta$ and $\kappa$ are shape and scale parameters of Gamma distribution of postsynaptic ISI obtained from the constraints $g_1 = \kappa/\beta$ and $g_0 = \psi(\kappa) - \log(\beta)$, and $\log(\cdot)$ and $\psi(\cdot)$ are natural logarithm and digamma functions.

The marginal distribution of postsynaptic ISI is computed as follows

$$f_{T|S}(\tau|s) = \int d\lambda_{exc} f_{\Lambda_{exc}|S}(\lambda_{exc}|s) f_{T|\Lambda_{exc},S}(\tau|\lambda_{exc},s). \tag{20}$$

Replacing (15) in (20), we have

$$\begin{aligned} f_{T|S}(\tau|s) &= \int d\lambda_{exc} f_{\Lambda_{exc}|S}(\lambda_{exc}|s)\,\mathbb{GAM}\left(b^{(\text{gam})}(s,\lambda_{exc}), m^{(\text{gam})}(s,\lambda_{exc})\right) \\ &= \frac{\beta^{\kappa}\tau^{\kappa-1}e^{-\beta\tau}}{\Gamma(\kappa)}u(\tau). \end{aligned} \tag{21}$$

Ideally, the above integration equation could be solved to obtain the distribution $f_{\Lambda_{exc}|S}(\lambda_{exc}|s)$ over the input rates. For the fitted Gamma distribution, we should derive the marginal distribution over the input rates. The optimal distribution for $\lambda_{exc}$, $f_{\Lambda_{exc}|S}(\lambda_{exc}|s)$, which maximizes the cost function in (9), for $15 \leq s \leq 45$ is given by

$$f_{\Lambda_{exc}|S}(\lambda_{exc}|s) = \beta^{\kappa} b_1^{(\text{gam})}(s) \frac{\Gamma\left(m_0^{(\text{gam})}(s)\right)}{\Gamma(\kappa)\Gamma\left(m_0^{(\text{gam})}(s)-\kappa\right)} \frac{\left(\lambda_{exc} b_1^{(\text{gam})}(s) - \beta + b_0^{(\text{gam})}(s)\right)^{m_0^{(\text{gam})}(s)-\kappa-1}}{\left(\lambda_{exc} b_1^{(\text{gam})}(s) + b_0^{(\text{gam})}(s)\right)^{m_0^{(\text{gam})}(s)}} \times \tag{22}$$

$$u(\lambda_{exc} b_1^{(\text{gam})}(s) - \beta + b_0^{(\text{gam})}(s))$$

where $\Gamma(.)$ denotes the Gamma function. The proof for deriving (22) is based on using Laplace transform for solving the integral equation of (21) and using fitting parameters of (16-a) in the integral equation of (21) (details are available in Appendix B).

Figure 7A shows a representative result for the optimization problem in (9) ($f_{\Lambda_{exc}}(\lambda_{exc})$) for $s = 30\%, 40\%$ and $50\%$, $E(\tau|s) = 100$ msec and $E(\log(\tau)|s) = -3.51$. It is evident that the minimal firing rate is about 25 Hz. Interestingly for higher synchronization levels, the minimal firing rate increases (Figure 7A). In the next subsection, we will use these optimal distributions of the input rates (Figure 7C) to determine the transmitted information in bits per unit cost, $\mathcal{I}_{bpj}$, and the associated synchronization level which maximizes $\mathcal{I}_{bpj}$ (optimization problem in (8)).

### C. Dependence of of Energy Efficiency and Synchronization Level

How does the information in bits per unit cost, $\mathcal{I}_{bpj}$, depend on the synchronization level (the control parameter)? To compute $\mathcal{I}_{bpj}$, we first need to compute the conditional entropy of the postsynaptic ISI, $h\left(T | \Lambda_{exc} = \lambda_{exc}, \Lambda_{inh} = \lambda_{inh}, S = s\right)$, which is shown in Figures 7B for different values of $\lambda_{exc}$ and $\lambda_{inh}$. The conditional entropy was estimated directly from the simulated postsynaptic ISIs, i.e., they do not depend on the choice of the parametric distribution used to fit the postsynaptic ISIs. As it can be seen in Figures 7B, for each synchronization level, the conditional entropy has a minimum at a particular input rate, $\lambda_{exc}$, whose value depends on the synchronization level (especially note the synchronization levels below 50% and that the minima shows the channel is less noisy).

Combining the conditional entropy estimates (Figure 7B) and the derived optimal input distribution (Eq. (22)), we have

$$\mathcal{I}_{bpj} = \frac{\mathcal{I}}{C_0 + C_1 E(T)},$$
$$= \frac{\mathcal{I}}{C_0 + C_1 \kappa/\beta}, \tag{23}$$

where $\mathcal{I}$ is defined in (11) and can be obtained by

$$\mathcal{I} = \mathcal{I}\left(\Lambda_{exc}; T | s\right) + \kappa E\left(\log T | s\right) - C$$
$$= h(T|s) + (\kappa - 1) E(\log T | s) - h\left(\log(T) | \log\left(d_1^{(b_{gam})}(s)\Lambda_{exc} + d_0^{(b_{gam})}(s)\right), s\right) - C,$$
$$= h(T|s) + (\kappa - 1) E(\log T | s) - h(\Omega) - C, \tag{24}$$

and $C$ is a constant value, and $\Omega$ is the transmission noise term. The $\Omega$ is independent from $\lambda_{exc}$, but is affected by the synchronization level, which in turn can modulate $\mathcal{I}_{bpj}$. Steps to prove this and how $\Omega$ depends on the synchronization level are based on using the fitting parameter of (16-a) in (24) (details are available in Appendix C).

Figure 7C shows $\mathcal{I}_{bpj}$ for different synchronization levels, where it can be seen that $\mathcal{I}_{bpj}$ is maximized at a non-zero synchronization level around 30%, in our model setup. Importantly, this finding indicates that i) synchronization can effectively modulate the energy efficiency of neuronal information transmission, and ii) the synchronized activity of just a portion of the population suffices for enabling an energy efficient transmission of presynaptic inputs to postsynaptic side.

Energy expenditure function, (10), includes two terms of $C_0$ and $C_1$. The $C_0$ term is related to the required energy for generating spike, and $C_1$ is related to postsynaptic neurons' energy expenditure per unit time in the interval of two successive spikes. Constraints on the energy expenditure function is given by

$$E\big(e(T)\big|S\big) < A \Rightarrow E\big(T\big|S\big) < \frac{A - C_0}{C_1} = B. \tag{25}$$

where we know that $C_0 > 0$, $C_1 > 0$, $A - C_0 > 0$. By increasing $C_1$ and $C_0$, $B$ decreases. As evident in Figure 8A, decreasing $B$, reduces $\mathcal{I}_{bpj}$ for fixed values of $E(\log(T))$. In other words, a tighter constraint on energy expenditure reduces $\mathcal{I}_{bpj}$. Note the synchronization level that maximizes $\mathcal{I}_{bpj}$ does not change by modifying the constraint value, i.e., $E(T|S)$. The maximum value of $\mathcal{I}_{bpj}$, $\mathcal{I}_{bpj}^{\max}$, depends on the conditional PDF of postsynaptic ISIs, i.e., $f_T\big(\tau\big|\lambda_{exc}, \lambda_{inh}, s\big)$ in (18). We also computed $\mathcal{I}_{bpj}$ in terms of synchronization level for different values of $E(\log(T))$, when $E(T)$ is fixed (Figure 8B). $E(\log(T))$ determines the mutual information between two successive spikes. Similar to Figure 8A, we found that the synchronization level that maximizes $\mathcal{I}_{bpj}$ does not change by modifying the constraint, which in this case (Figure 8B), is regulated by $E(\log(T))$. Collectively, these results (Figure 8A and 8B) imply that the synchronization level maximizing $\mathcal{I}_{bpj}$ does not change by constraints on $E(T)$ and $E(\log(T))$. Instead, in our model this is controlled by $f_T\big(\tau\big|\lambda_{exc}, \lambda_{inh}, s\big)$.

To study how the correlation between two successive spikes affects $\mathcal{I}_{bpj}^{max}$, we computed the maximum value of $\mathcal{I}_{bpj}$ in terms of mutual information between $T_1$ and $T_2$ in (9) (Figure 8C). To this end, we found $\kappa$ from optimization problem (9), and set $C = 8.6696$ so that the positivity of $\mathcal{I}(T_1; T_2)$ can be guaranteed. It can be seen that by increasing $\mathcal{I}(T_1; T_2)$, $\mathcal{I}_{bpj}^{max}$ first increases and then decreases. As such, when $\mathcal{I}(T_1; T_2)$ increases beyond a certain value, $\mathcal{I}_{bpj}^{max}$ is reduced (Figure 8C).

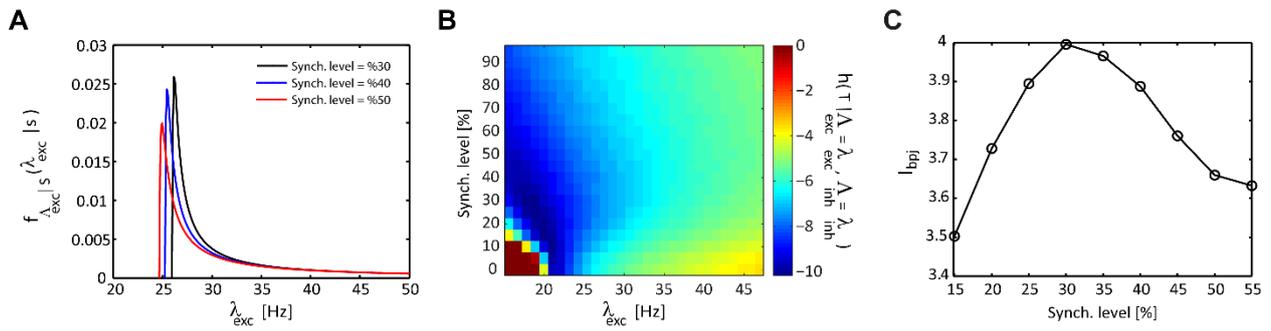

**Figure 7.** Optimal input distribution and the resulting entropy and information content. *A,* optimal distribution of presynaptic excitatory firing rate (input rate; $\lambda_{exc}$) given synchronization level, $f_{\Lambda_{exc}|S}(\lambda_{exc}|s)$. The result is derived by solving optimization problem in (9) for different values of *s*. *B,* conditional entropy of postsynaptic ISIs, $h(T|\Lambda_{exc} = \lambda_{exc}, \Lambda_{inh} = \lambda_{inh}, S = s)$, in terms of $\lambda_{exc}$ and synchronization levels, $s$. The firing rate of presynaptic inhibitory neurons, $\lambda_{inh}$, was set to 125 sp/s. *C,* the values of information per unit cost, $\mathcal{I}_{bpj}$, for the optimal input distribution (for $C_0 = 1$).

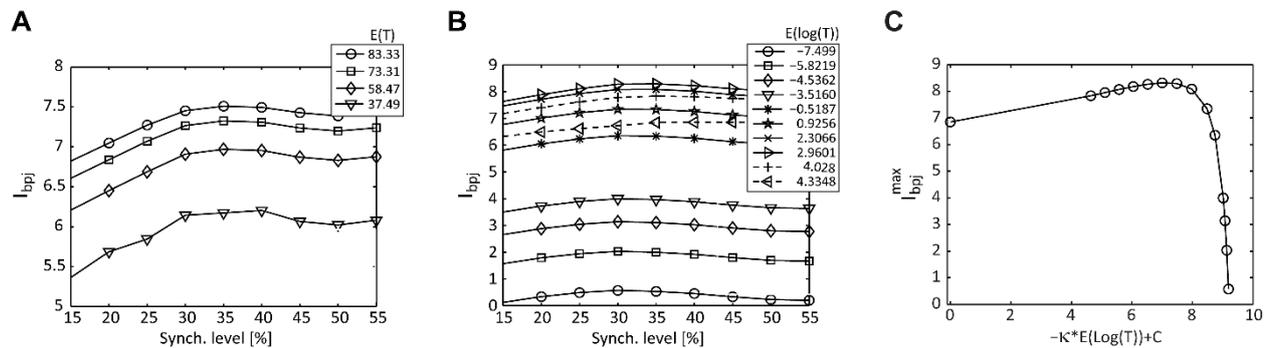

**Figure 8.** Average mutual information per unit cost, $\mathcal{I}_{bpj}$, as a function of synchronization level. *A,B,* $\mathcal{I}_{bpj}$ in terms of synchronization level for: *A,* different values of expected values of postsynaptic ISIs, $E(T)$,

and $E(\log(T)) = -5$, **B**, different values of $E(\log(T))$ and $E(T) = 100$ *m*sec . **C**, the maximum value of $\mathcal{I}_{bpj}$ vs. the mutual information between successive postsynaptic ISI, $\mathcal{I}(T_1;T_2)$, for $E(T) = 100$ *m*sec .

## IV. DISCUSSIONS

We studied the role of neuronal synchronization as an independent control parameter in a population of presynaptic excitatory neurons (as a transmitter) on the information coding of a biologically realistic postsynaptic neuron model (as a receiver). This is a new way of conceptualizing synchronicity compared to previous studies. We showed that the synchronization can effectively influence the postsynaptic ISI distribution and therefore – within the Berger-Levy theory– also the energy-efficiency of neural communication. More specifically, for a given input rate, we found that the conditional distribution of postsynaptic ISI is well-described by Gamma and generalized extreme value (GEV) distributions for synchronization levels of 0-50% and 0-100%, respectively. Using a Gamma distribution, we computed the optimal input distribution which maximizes the mutual information per unit cost, for synchronization levels of 15% to 50% (restricted for technical reasons). Our analysis showed that there is a maximum at approx. 30% synchronization level for the mutual information per unit cost. This suggests that a synchronized activity of just a certain portion of the population (i.e. ~30%) is a signature of optimal energy efficient transmission of sensory inputs. In our analysis, this energy efficient transmission was obtained through minimization of the transmission noise term. These results may aid in better understanding of how neuronal synchronization modulate and facilitate information transmission in neuronal communications.

### A. *Effect of Neuronal Synchronization on Information Transmission during Attention*

Synchronization is a ubiquitous phenomenon in neuronal networks, and has been observed at different structural ranges: i) short-range synchronization between individual neurons of a cortical microcircuit, and ii) long-range synchronization between cortical regions. Several studies have already reported the efficiency of synchronization (among the presynaptic excitatory neurons) for neuronal coding [76-78], and investigated its potential roles in cognitive functions such as memory, learning, attention, perception, and development [45, 78-81]. We established our energy efficiency study based on the role of short-range (or local) synchronization in attention, which plays a central role in cognitive processing [45, 82]. Nevertheless, our main findings may also be able to explain the communicational role of synchronization in other brain functions as well (see also below). In this line, our findings as reflected in, e.g. Figure 7C, can potentially be used in future works to identify, and thus better understand, the biological mechanisms and/or physiological parameters which play a critical role in governing the optimal energy efficiency of neuronal communication, in the underlying networks.

For synchronization levels between 0% (asynchronous case) to 100% (whole-population synchronization case), we found in both simulation scenarios of synaptic inputs modelling, that the conditional entropy of the postsynaptic ISI for given values of $\lambda_{exc}$ has a global minimum (see Figure 7C) at the synchronization level around 30%. Based on the potential role of synchronization, e.g., in attention [45], this finding implies that attention tends to make information transmission more efficient (through minimization of the conditional entropy of postsynaptic ISI for the given $\lambda_{exc}$). More generally, while $\lambda_{exc}$ of individual neurons may encode the stimulus features, their synchronized firing patterns can effectively regulate the energy efficiency of the ISI coding used by postsynaptic side to transmit the information about the presynaptic average firing activity.

*B. Role of Neuronal Synchronization in Modulating Firing Characteristics of Postsynaptic Neuron*

We quantified the potential effects of the synchronization on the firing charachteristics of the postsynaptic neuron. To this end, we used two simulation scenarios for modeling the balanced regimes of excitatory and inhibitory synaptic inputs (see Methods), and we systematically increased the synchronization level. We found that: i) in both scenarios, such increase does not affect the mean firing rate of the postsyanptic neuron (see, e.g., Figure 2A), ii) in the excitation-driven scenario, increasing the synchronization decreases the threshold of the ascending phase in the firing rate of postsynaptic neuron, iii) in both scenarios, the CV of postsynaptic ISIs increases at higher synchronization levels (see, e.g., Figure 2B).

Furthermore, we studied the role of synchronization in shaping the postsynaptic ISI distribution, by incorporating the synchronization as an additional independent parameter into our simulations and information theory framework. To our knowledge, the effect of synchronization had been neglected by previous studies [24, 34-36]. For instance, while [24] studied the efficiency of neuronal communication under a Gamma distribution assumption for the postsynaptic ISI distribution (for given $\lambda_{exc}$), our results indicate that such assumption may not be allowed if synchronization is taken into consideration. This is because we found that for synchronization levels bigger than approximately 50%, Gamma distribution does not fit well to the postsyanptic ISI distribution, for the given $\lambda_{exc}$ (Figure 4B). We showed that the GEV distribution can, instead, provide a reliable fit to this distribution for all synchronization levels.

### C. Implications of Neuronal Synchronization for Energy Efficiency of Neuronal Coding

We found that the average mutual information per unit cost is maximized at the synchronization level of about 30% (Figure 7C). This finding implies that the synchronization between just a portion of excitatory neurons of the corresponding population may be sufficient to allow for an energy efficient information transmission during neural computations; e.g. during processing of sensory inputs. This optimal information transmission will increase the energy efficiency of the postsynaptic neuron (in terms of $\mathcal{I}_{bpj}$). Strikingly, such a synchronization level of approximately 30% is in accordance with the exprimental observations during stimulus-evoked and spontaneous Up-state activities in mature neuronal networks [39-41, 83]. The combination of these exprimental and our theoritical findings suggests that mature neuronal networks, by utilizing some particular relatively low levels of synchronization, may aim at optimizing the energy expenditure of their inforamtion processing.

Unlike adult networks, eliciting highly synchronized activitity patterns is a ubiquitous property of many neonatal neural structures [43, 84-86]. However, during neocortical development, the activity of developing neural networks transitions from this highly synchronized mode (with a synchronization level of about 80%) to a relatively sparse mode after the postnatal onset of sensory transduction [44, 87-89]. While this so-called sparsification phenomenon is thought to refine sensory coding [58], our findings may provide some additional mechanistic explanation for the actual purpose of sparsification. That is, our resutls (Figure 7C) suggest that the neuronal communication in immature cortices lacking sensory inputs (e.g., in mice visual cortex during first postnatal week [43, 44, 90]) may not be reliably energy efficient. However, the sparsification refines the network activity such that the synchronization can be used in an optimized fashion to make the neuronal information processing more energy efficient.

It has been found that abnormally strong; pathological neuronal synchronization can be a hallmark of several neurological disorders such as epilepsy, tinnitus, essential tremor, and Parkinson's disease (see [91] for a review). This follows as, under normal physiological conditions, the neurons of the same networks exhibit an asynchronous state or a relatively small level of synchronization. Accordingly, our approach, by computing the energy efficiency of neuronal communication at different synchronization levels, provides novel insights into the effects of excessive level of synchronization in these diseases. In motor impairment [92]), it is reported that the abnormally strong synchronization precludes the underlying networks from processing information in an energy efficient manner. We found that the average mutual information per unit cost is maximized at a relatively small synchronization level (~30%, Figure 7C) and drops quickly as synchronization level increases. Moreover, this finding may aid in the calibration of the desynchronizing stimulation technique, used in deep brain stimulation for the treatment of the abovementioned diseases [93]. Specifically, it may guide the therapists to perform the desynchronization stimulation such that, instead of shifting the activity of underlying network to a relatively asynchronous state (anti-kindling process) [91, 93], push it to a mode with a relatively small synchronization level so that the network's energy efficient state can be provoked. This may, in turn, provide a better treatment procedure.

*D. Limitations and Extensions*

In this work, to establish our information theoretic framework we aimed at using less sophisticated, but still biologically plausible, models. However, there is some room for improvement. For instance, to model the synaptic inputs and input spike trains we followed previous, well-established modeling studies [37, 94]. However, the synaptic input model lacks, and thus can be extended to incorporate, several detailed mechanisms of synaptic transmission. These include the vesicle release machinery [14, 95], short-term synaptic plasticity [96], and the neurotransmitters' reuptake and diffusion [97], or the axonal and synaptic noise [98, 99]. The simple (homogenous) Poisson process, which we used for modelling the input spike trains (see also [37, 66]), can also be replaced by e.g. a Gamma process [100, 101] or a doubly Poisson process [98]. In general, these extensions can provide a more accurate approximation of the synaptic transmission, and the firing statistics of input spike trains. Moreover, to achieve a more realistic emulation of neuronal synchronizations [42, 86, 102], the zero-time-lag synchronous spike patterns used in this work can be jittered within a plausible specified time window. Taken together, future works are required to evaluate to what extent our main findings (e.g. see Fig. 7C) are affected by such potential extensions and their corresponding parameterizations, or by using e.g. i) other types of postsynaptic neuron models than the HH-type we used here (e.g. with different excitability type, or with bursting behavior [103]), or ii) different synaptic excitation to inhibition ratios (E/I), or iii) presynaptic neurons with heterogeneous firing characteristics.

Furthermore, as we showed in this work, for synchronization levels below 50%, the bit per joule maximization problem (Eq. (9)) can be solved analytically (Eq. (22)), when considering the Gamma distribution as the conditional probability of neuronal communication channel. However, the GEV distribution fits better to the conditional probability of postsynaptic ISI, for given $\lambda_{exc}$, at all synchronization levels. Therefore, finding an analytical solution for the maximization problem with GEV distribution may provide a better understanding of the synchronization role in modulating the average mutual information per unit cost over neuronal communications.

# Supplementary Material

We obtain ISI distributions by simulating a Hodgkin-Huxley-type model [1, 2] neuron using two defined simulation scenarios. In the excitation-driven simulation scenario, which is described in the paper, we keep the inhibitory rate $\lambda_{inh}$ constant. Therefore, for brevity, we now drop $\lambda_{inh}$ in $f(t|\lambda_{exc}, \lambda_{inh}, s)$ and denote it with $f(t|\lambda_{exc}, s)$. In the fluctuations-driven simulation scenario, where the results of that are reported here, we co-vary the inhibitory rate $\lambda_{inh}$ together with the excitatory input $\lambda_{exc}$. The rational for this choice was the two scenarios correspond to two biophysically different transmission regimes (both of them having strong excitation *and* inhibition). In the excitation-driven simulation scenario, the responses are due to excitatory drive whereas in the fluctuations-driven simulation scenario the responses are exclusively due to changes in the fluctuations causes by a balanced change of $\lambda_{exc}$ and $\lambda_{inh}$. In terms of the system model, the constant inhibition corresponds to viewing $\lambda_{inh}$ also as a control parameter whereas in the fluctuations-simulation scenario, the inhibition $\lambda_{inh}$ is also part of the input (with a strict dependence on $\lambda_{exc}$). In both scenarios, however, the synchronicity *S* among the excitatory neurons is the channel's control parameter we are primarily interested in. Here, the simulation results of fluctuations-driven scenario is provided. Also, some technical results from fluctuations-driven simulation scenario, which are refered in the paper are reported. Finally details of proof for proposed arguments, which the proof of concept of them are proposed in the paper, are proposed in the supplementary materials.

## SI- FLUCTUATIONS-DRIVEN SIMULATION SCENARIO

Here, we consider the fluctuations-driven simulation scenario, in which both $\lambda_{exc}$ and $\lambda_{inh}$ are varied. We enforce the synchronization to the different fraction of presynaptic neurons. We have simulated the postsynaptic neuron for different levels of the presynaptic neurons' synchronization (0%, 30% and 90%) . We applied the synchronization only to the excitatory neurons. In Figure S1A, B and D the ratio of $G_{exc}$ (sum conductance of all presyanaptic excitatory neurons) to $g_L$ (leakage conductance ) has shown. There can be seen, in each of these panels, the higher levels of synchronization leads to increasing the standard deviation, of the $G_{exc}/G_{leakage}$, but their time-averaged remains same. The difference between these panels is in the time-averaged conductance related to each level firing rate. Indeed, for higher levels of presynaptic firing rates the $G_{exc}/g_L$ is elevated. The reason of this behavior originates in the essence of the Poisson process, which have beed employed to generate the presynaptic spike trains. Figure S1.C illustrates the ratio of $G_{inh}$ (summed conductance of all presynaptic inhibitory neurons) to $g_L$ for different levels of $\lambda_{exc}$, where there is no synchronization between the inhibitory ones. It should be noted that the synchronization in presynaptic excitatory neurons does have no effect on the inhibitory ones and vise versa, because the spike trains of these population (and all synapses) are independent.

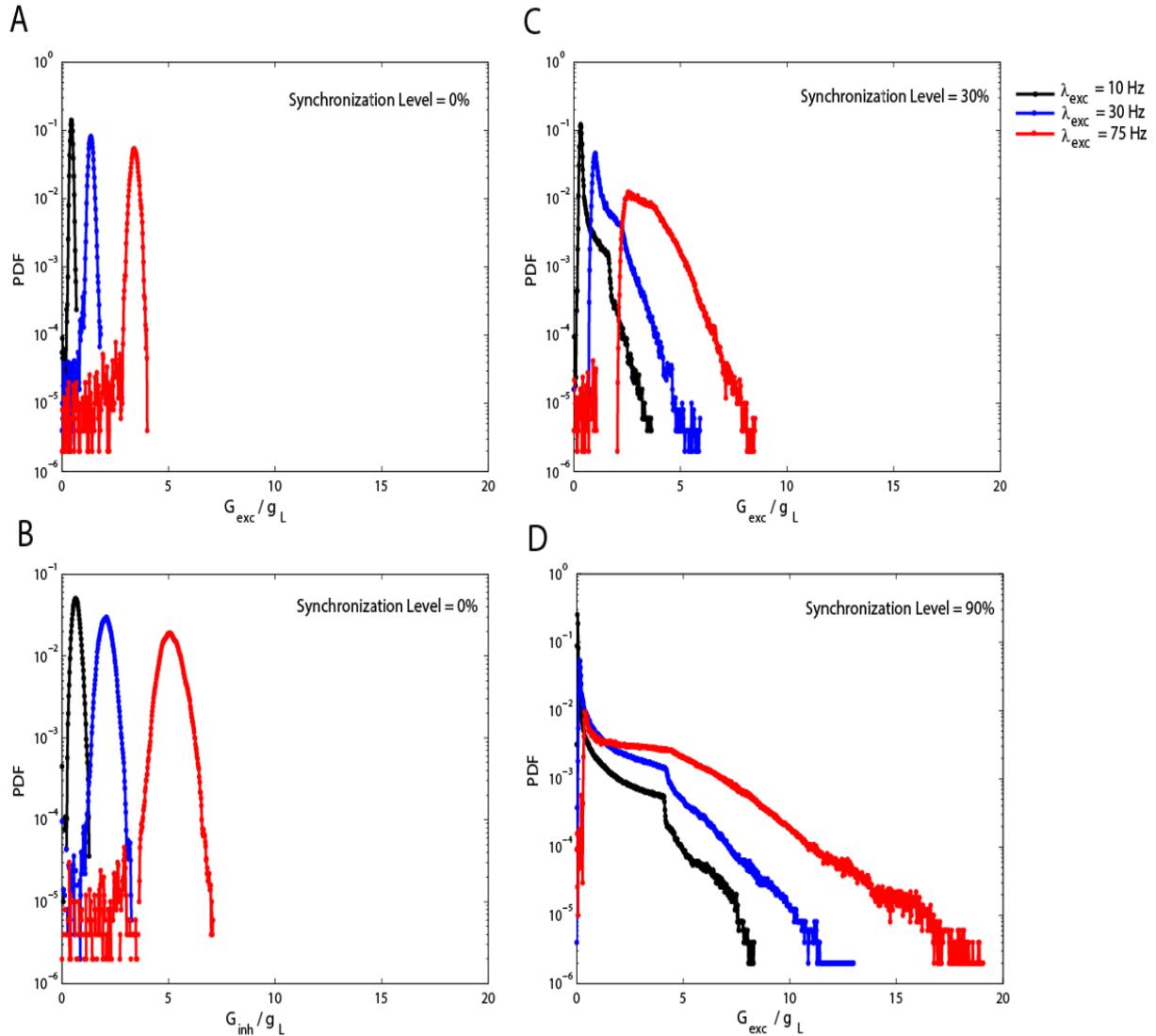

**Figure S1.** PDF of $G_{exc}/g_L$ and $G_{inh}/g_L$ for different values of $\lambda_{exc}$ and synchronization level. *A,* PDF of $G_{exc}/g_L$ for different values of $\lambda_{exc}$ with synchronization level of 0%. *B,* PDF of $G_{inh}/g_L$ for different values of $\lambda_{exc}$ with synchronization level of 0%. *C,* PDF of $G_{exc}/g_L$ for different values of $\lambda_{exc}$ with synchronization level of 30%. *D,* PDF of $G_{exc}/g_L$ for different values of $\lambda_{exc}$ with synchronization level of 90%.

Figure S2A shows PDF of $V_{sub}$ (sub-threshold voltage) in mV for different values of $\lambda_{exc}$ with synchronization level of 0%. It can be observed in this sub-figure, $V_{sub}$ has almost similar shape for different values of $\lambda_{exc}$. Figure S2B shows PDF of $V_{sub}$ in mV for different values of $\lambda_{exc}$ with synchronization level of 30%, here it can be observed that by increasing synchronization level to 30%, smaller values of sub-threshold voltage has more probability. Figure S2C shows PDF of $V_{sub}$ in mV for different values of $\lambda_{exc}$ with synchronization level of 30%, it can be observed that by increasing synchronization level to 90%, the shape of PDF for different values of $\lambda_{exc}$ are changed and smaller values of $V_{sub}$ has more probability.

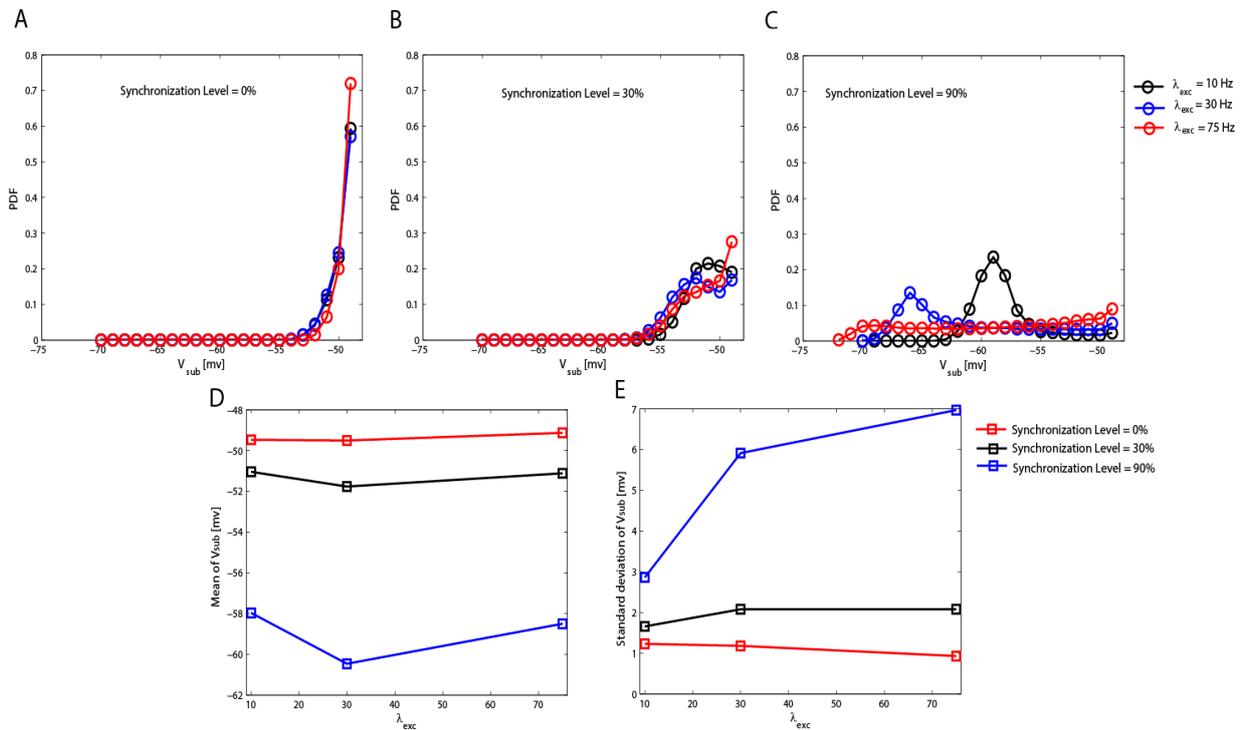

**Figure S2.** Different Statstics of $V_{sub}$ (sub-threshold voltage) in mV in terms of $\lambda_{exc}$ for different values of sysnchronization level. **A,** PDF of $V_{sub}$ in mV for different values of $\lambda_{exc}$ with synchronization level of 0%. **B,** PDF of $V$ in mV for different values of $\lambda_{exc}$ with synchronization level of 30%. **C,** PDF of $V$ in mV for different values of $\lambda_{exc}$ with synchronization level of 90%. **D,** Mean of subthereshold voltage in terms of $\lambda_{exc}$ for different level of synchronization level. **E,** Standard deviation of subthereshold voltage in terms of $\lambda_{exc}$ for different level of synchronization level. in all subfigures subthereshold voltage is -48 mV.

Figure S2D shows the mean of $V_{sub}$, in terms of $\lambda_{exc}$ for different values of synchronization levels, it is demonstrated that mean of $V_{sub}$ by increasing $\lambda_{exc}$ is reduced. Figure S2E shows the standard deviation of of $V_{sub}$, in terms of $\lambda_{exc}$ for different values of synchronization levels, it is demonstrated that mean of $V_{sub}$ by increasing $\lambda_{exc}$ is increased, which is matched with our expectation from increasing fluctuations of $V_{sub}$ due to increasing synchronization level.

## SII- CHARACTERSTICS OF PARAMETRIC DISTRIBUTIONS

Table S1 summarizes the distributions, their parameters, and their maximum likelihood estimates.

Table S1. Characteristics of parametric distributions [45].

| Distribution | Parameters | | | PDF - $f_T(\tau)$ | Maximum likelihood estimate of PDF parameters |
|---|---|---|---|---|---|
| | Scale | Shape | Location | | |
| **Exponential** | $\mu^{(exp)} > 0$ (mean) | - | - | $\mathbb{EXP}(\mu^{(exp)}) = \frac{1}{\mu^{(exp)}} e^{-\frac{\tau}{\mu^{(exp)}}}$ | $\hat{\mu}^{(exp)} = \bar{\tau} = \frac{1}{n}\sum_{i=1}^{n} \tau_i$ |
| **Gaussian** | $\sigma^{(gau)} > 0$, (standard deviation) | - | $\mu^{(gau)} \in \mathbb{R}$ (mean) | $\mathbb{GAU}(\sigma^{(gau)}, \mu^{(gau)}) = \frac{1}{\sigma^{(gau)}\sqrt{2\pi}} e^{-\frac{(\tau-\mu^{(gau)})^2}{2(\sigma^{(gau)})^2}}$ | $\hat{\mu}^{(gau)} = \bar{\tau} = \frac{1}{n}\sum_{i=1}^{n} \tau_i$, $\hat{\sigma}^{(gau)} = \sqrt{\frac{1}{n-1}\sum_{i=1}^{n}(\tau_i - \bar{\tau})^2}$ |
| **Weibul** | $\theta^{(wei)} > 0$ | $m^{(wei)} > 0$ | - | $\mathbb{WEI}(\theta^{(wei)}, m^{(wei)}) = \left(\frac{m^{(wei)}}{\theta^{(wei)}}\right)\left(\frac{\tau}{\theta^{(wei)}}\right)^{m_{wei}-1} \times e^{-(\tau/\theta^{(wei)})^{m^{(wei)}}} u(\tau)$ | $\hat{\theta}_{wei}^{m^{(wei)}} = \frac{1}{n}\sum_{i=1}^{n}\left(\tau_i^{m^{(wei)}} - \tau_n^{m^{(wei)}}\right)$ where $\tau_1 > \tau_2 > ... > \tau_n$ are the $n$ largest observed samples $\frac{1}{\hat{m}^{(wei)}} = \frac{\sum_{i=1}^{n}\left(\tau_i^{m^{(wei)}} \ln \tau_i - \tau_n^{m^{(wei)}} \ln \tau_n\right)}{\sum_{i=1}^{n}\left(\tau_i^{m^{(wei)}} - \tau_n^{m^{(wei)}}\right)} - \frac{1}{n}\sum_{i=1}^{n} \ln \tau_i$ |
| **Gamma** | $\frac{1}{b^{(gam)}} > 0$ | $m^{(gam)} > 0$ | - | $\mathbb{GAM}(b^{(gam)}, m^{(gam)}) = \frac{(b^{(gam)})^{m^{(gam)}} \tau^{m^{(gam)}-1} e^{-b^{(gam)}\tau}}{\Gamma(m^{(gam)})} u(\tau)$ | $\hat{b}^{(gam)} = \frac{1}{\frac{1}{nm^{(gam)}}\sum_{i=1}^{n} \tau_i}$ $m^{(gam)} \approx \frac{3-\ell+\sqrt{(\ell-3)^2+24\ell}}{12\ell}$ $m^{(gam)} \leftarrow m^{(gam)} - \frac{\ln(m^{(gam)}) - \psi(m^{(gam)}) - \ell}{1/m^{(gam)} - \psi'(m^{(gam)})}$, $\ell = \ln\left(\frac{1}{n}\sum_{i=1}^{n} \tau_i\right) - \frac{1}{n}\sum_{i=1}^{n} \ln(\tau_i)$ |
| **Log Normal** | $\mu^{(\log)} \in \mathbb{R}$ (log scale) | $\sigma^{(\log)} > 0$ | - | $\mathbb{LOG}\left(\mu^{(\log)}, \sigma^{(\log)}\right) = \dfrac{1}{\tau\sqrt{2\pi}\sigma^{(\log)}} e^{\dfrac{-\left(\ln\tau - \mu^{(\log)}\right)^2}{2\left(\sigma^{(\log)}\right)^2}}$ | $\hat{\mu}^{(\log)} = \dfrac{1}{n}\sum_{i=1}^{n} \ln\tau_i$ $\hat{\sigma}^{(\log)} = \sqrt{\dfrac{1}{n}\sum_{i=1}^{n}\left(\ln\tau_i - \hat{\mu}^{(\log)}\right)}$ |
| **Generalized Extreme Value** | $\mu^{(gev)} \in \mathbb{R}$ | $\sigma^{(gev)} > 0$ | $k^{(gev)} \in \mathbb{R}$ | $\mathbb{GEV}\left(\mu^{(gev)}, \sigma^{(gev)}, k^{(gev)}\right) = \dfrac{1}{\sigma^{(gev)}} \vartheta(\tau)^{k^{(gev)}+1} e^{-\vartheta(\tau)}$ $\vartheta(\tau) = \begin{cases}\left(1 + \left(\dfrac{\tau - \mu^{(gev)}}{\sigma^{(gev)}}\right)k^{(gev)}\right)^{-1/k^{(gev)}}, k^{(gev)} \neq 0 \\ e^{-\left(\tau - \mu^{(gev)}\right)/\sigma^{(gev)}}, k^{(gev)} = 0\end{cases}$ | Based on an iterative algorithm presented in [46] |

SIII-MORE RESULTS ON EXCITATION-DRIVEN SIMULATION SCENARIO

The result of Kolmogorov-Smirnov(KS) test ($q = 1(0)$ corresponds to rejection(acceptance)), and corresponding p-values are shown, respectively, in Figure S3 and S4 for Gamma, Gaussian, Log-normal, Weibull, Exponantial and Generalized extreme value distributions in terms of synchronization level and $\lambda_{exc}$. In Figure S3 the blue region shows the area (corresponding to values of $\lambda_{exc}$ and $S$), in which KS test are acceptable. While in Figure S4, the blue area shows smaller p-values, which means that KS test is not acceptable.

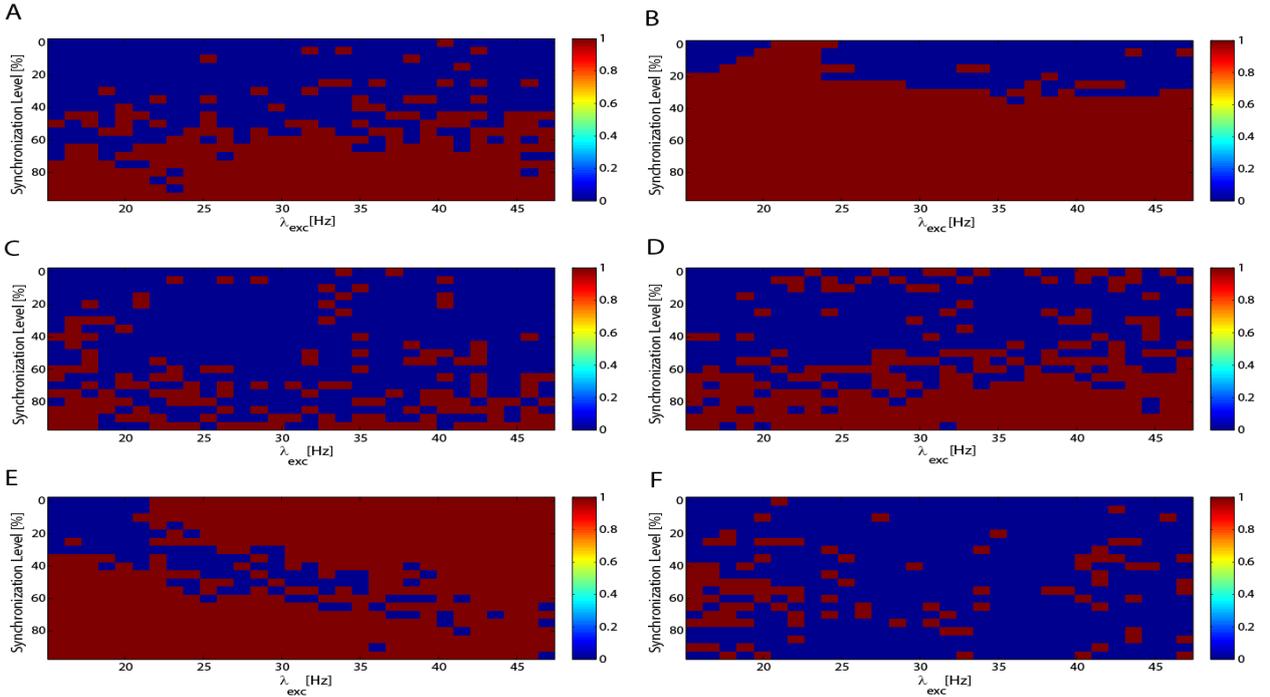

**Figure S3.** The result of Kolmogorov-Smirnov test ($q = 1(0)$ corresponds to rejection(acceptance) of test) for **A**, Gamma **B**, Gaussian **C**, Log-normal **D**, Weibull **E**, Exponantial and **F**, Generalized extreme value distributions in terms of synchronization level and $\lambda_{exc}$.

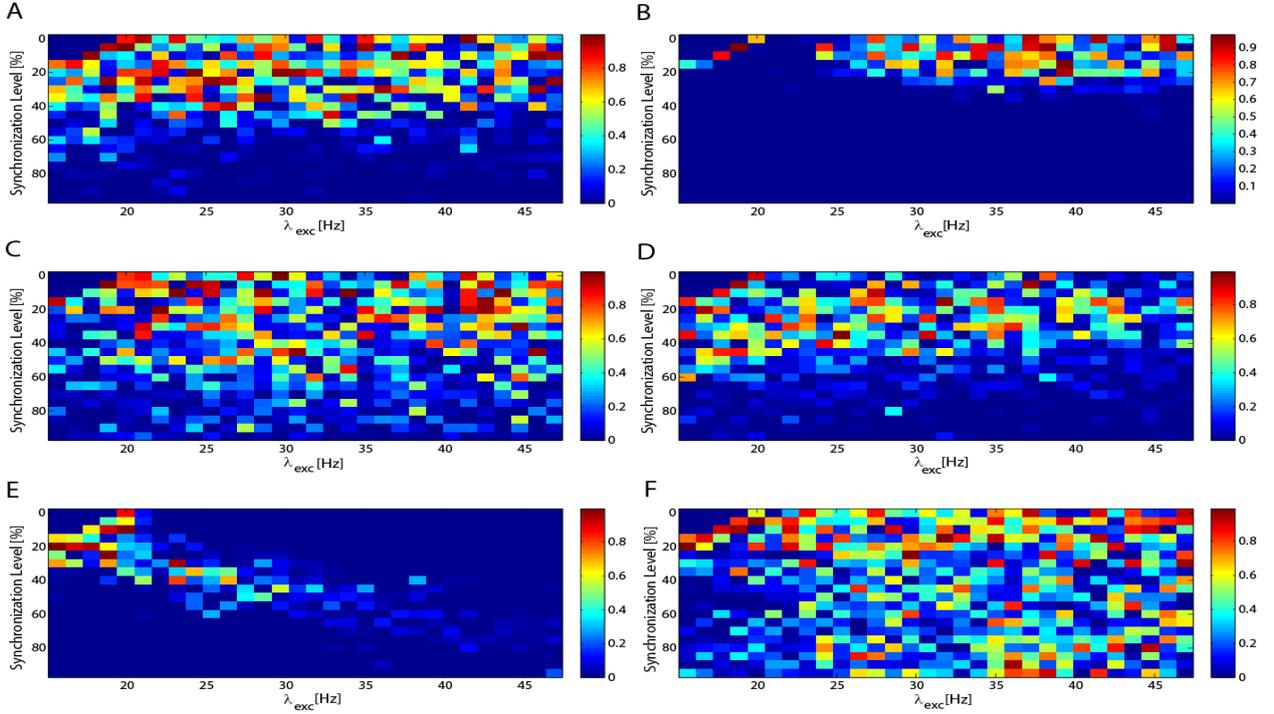

**Figure S4.** p-value of Kolmogorov-Smirnov test for **A**, Gamma **B**, Gaussian **C**, Log-normal **D**, Weibull **E**, Exponantial and **F**, Generalized extreme value distributions in terms of synchronization level and $\lambda_{exc}$.

Figure S5A-C show $k_i^{(gev)}(s)$, $\sigma_i^{(gev)}(s)$ and $\mu_i^{(gev)}(s)$, ($i \in \{1,2,3\}$), which are fitted to the parameters of Generalized extreme value distribution. Figure S6A and B show $m_i^{(gam)}(s)$, $i \in \{1,2,3\}$ and $b_i^{(gam)}(s)$, $i \in \{1,2\}$, which are fitted to the paraemetrs of Gamma distribution

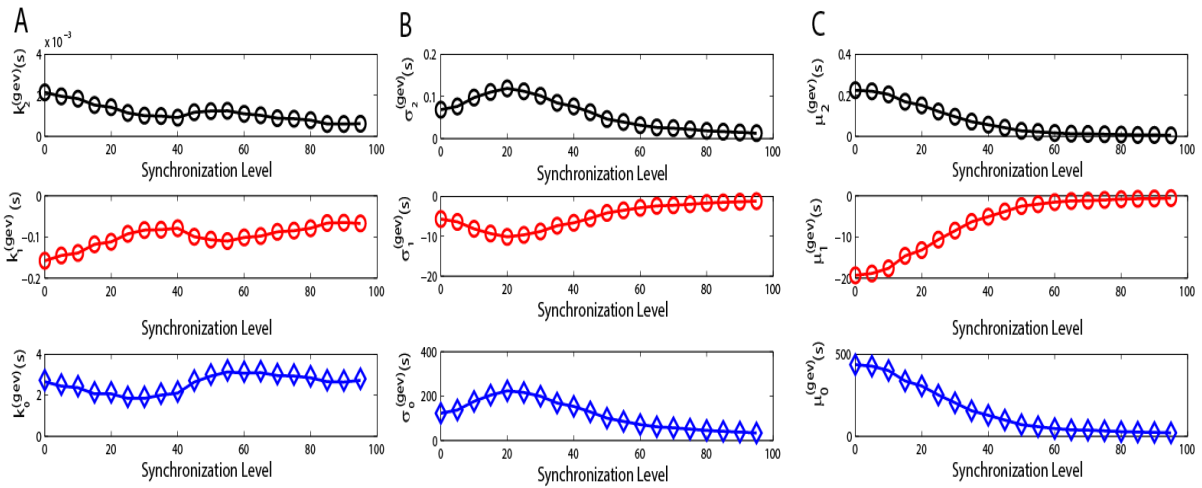

**Figure S5.** Coefficients of fitted quadratic functions to estimated parameters of GEV distribution. *A,* $k_i^{(gev)}(s)$ in terms of $s$. *B,* $\sigma_i^{(gev)}(s)$ in terms of $s$. *C,* $\mu_i^{(gev)}(s)$ in terms of $s$ with $i \in \{1,2,3\}$.

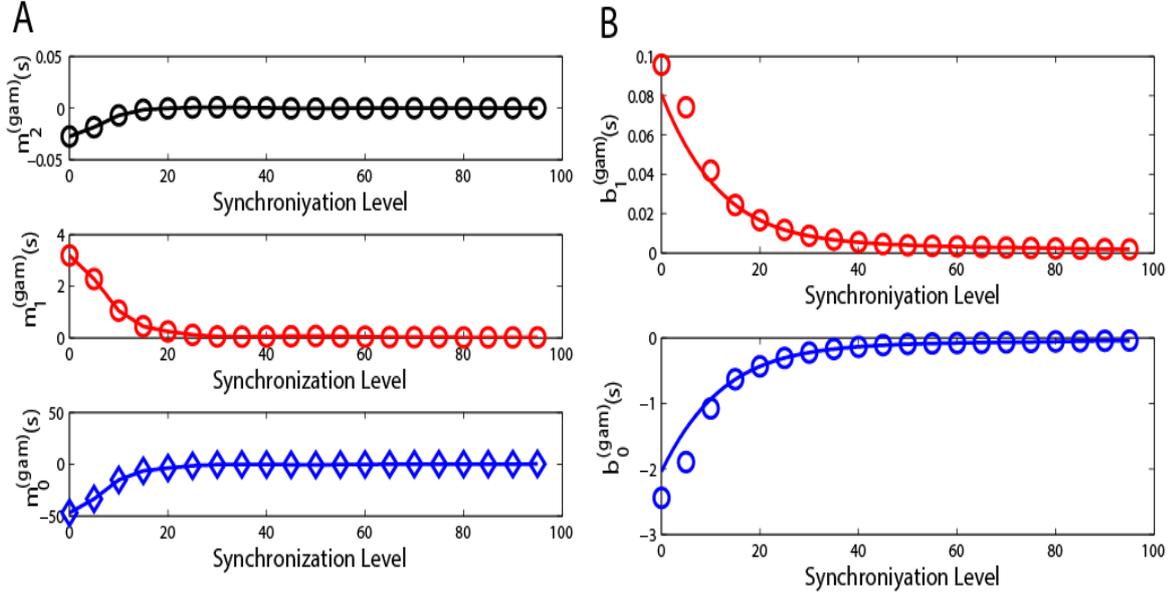

**Figure S6.** Coefficients of fitted functions to estimated parameters of Gamma distribution, *A,* $m_i^{(gam)}(s)$, $i \in \{0,1,2\}$ in terms of synchronization level. *B,* $b_i^{(gam)}(s)$, $i \in \{0,1\}$ in terms of synchronization level.

## SIV-APPENDICES

### A. Appendix A

The BPJ optimization problem as follows

$$I_{bpj}(s) = \max_{F_{\Lambda_{exc}|S}(\lambda_{exc}|S=s)} \frac{I}{E(e(t)|S=s)},$$

s.t.

$$F_{\Lambda_{exc}|S}(\lambda_{exc}|S=s) = \Pr(\Lambda_{exc} < \lambda_{exc}|S=s),$$

(S1)

is equivalent to following optimization problem

$$\max_{F_{T|S}(t|s)} h(T|s),$$ (S2-a)

s.t.

$$F_{T|S}(t|s) = \Pr(T<t|s),$$ (S2-b)

$$E(T|s) = g_0,$$ (S2-c)

$$E(\log_e T|s) = g_1,$$ (S2-d)

**Proof.** The proof begins by considering a Lagrangian function of (14). Similar to [3], the Lagrange function is rewritten as

$$J\left(F_{\Lambda_{exc}|S}\right) := I(\Lambda_{exc};T|s) - I(T_1;T_2|s) - \alpha' E\left[e(T)|s\right] - \beta'. \tag{S3}$$

where $\alpha'$ and $\beta'$ are Lagrange multipliers. Replacing $I(T_1;T_2|s) = -\kappa \log T + C$ [3, 4] and $e(T)$ from (13), $J\left(F_{\Lambda_{exc}|S}\right)$ is given by

$$J\left(F_{\Lambda_{exc}|S}\right) := I(\Lambda_{exc};T|s) + \kappa E\left[\log T|s\right] - \alpha'' E\left[T|s\right] - \beta'', \tag{S4}$$

where $\alpha'' = C_1 \alpha'$, $\beta'' = \beta + \alpha C_0 + C$ are constants. Also, the expectation $E$ is with respect to $T$, and $I(\Lambda_{exc};T|s)$ is given by

$$\begin{aligned}
I(\Lambda_{exc};T|s) &= I\left(d_1^{(b_{gam})}(s)\Lambda_{exc} + d_0^{(b_{gam})}(s); T|s\right) \\
&= I\left(\log\left(d_1^{(b_{gam})}(s)\Lambda_{exc} + d_0^{(b_{gam})}(s)\right); \log(T)|s\right) \\
&= h\left(\log(T)|s\right) - h\left(\log(T)|\log\left(d_1^{(b_{gam})}(s)\Lambda_{exc} + d_0^{(b_{gam})}(s)\right), s\right) \\
&\stackrel{(a)}{=} h\left(\log(T)|s\right) - h(\Omega) \\
&\stackrel{(b)}{=} -E\left(\log f_{Z|S}(Z|s)|s\right) - h(\Omega) \\
&\stackrel{(c)}{=} -E\left[\log\left(f_{T|S}(T|s).T\right)|s\right] - h(\Omega) \\
&\stackrel{(d)}{=} h(T|s) - E\left[\log(T)|s\right] - h(\Omega),
\end{aligned} \tag{S5}$$

where $\Omega$ interprets as noise terms.

(a) for synchronization level between 15% to 50%, where according to Figure S8 some of the fitting parameters approach zero, $f_{T|\Lambda_{exc},S}(t|\lambda_{exc},s)$ in (19) can be simplified to

$$f_{T|\Lambda_{exc},S}(t|\lambda_{exc},s) = \frac{\left(b_1^{(gam)}(s)\lambda_{exc} + b_0^{(gam)}(s)\right)^{m_0^{(gam)}(s)} t^{m_0^{(gam)}(s)-1} e^{-\left(b_1^{(gam)}(s)\lambda_{exc} + b_0^{(gam)}(s)\right)t}}{\Gamma\left(m_0^{(gam)}(s)\right)} u(t) \tag{S6}$$

By selecting $x = \left(b_1^{(gam)}(s)\lambda_{exc} + b_0^{(gam)}(s)\right)t$ we have the following equality

$$\left|f_{X|\Lambda_{exc},S}(x|\lambda_{exc},s)dx\right| = \left|f_{T|\Lambda_{exc},S}(t|\lambda_{exc},s)dt\right| \tag{S7}$$

It follows

$$f_{X|\Lambda_{exc},S}(x|\lambda_{exc},s) = f_{X|\Lambda_{exc},S}(x|s) = \frac{x^{m_0^{(gam)}(s)} e^{-xt}}{\Gamma\left(m_0^{(gam)}(s)\right)} u(t). \tag{S8}$$

$\lambda_{exc}$ does not appear in right hand side of the above equation and hence, $x$ is independent of $\lambda_{exc}$. This is despite the fact that $x = \left(b_1^{(gam)}(s)\lambda_{exc} + b_0^{(gam)}(s)\right)t$. We can rewrite the relationship as

$$t = \frac{1}{\left(b_1^{(\text{gam})}(s)\lambda_{exc} + b_0^{(\text{gam})}(s)\right)} x \tag{S9}$$

where $X$ is marginally distributed according to (S8). Then by taking logarithm in (S9) we have

$$\log t = -\log\left(b_1^{(\text{gam})}(s)\lambda_{exc} + b_0^{(\text{gam})}(s)\right) + \log x \tag{S10}$$

where $\log X = \Omega$ is independent of $\log\left(b_1^{(\text{gam})}(s)\lambda_{exc} + b_0^{(\text{gam})}(s)\right)$. Therefore, we have $h\left(\log(T)\big|\log\left(b_1^{(\text{gam})}(s)\lambda_{exc} + b_0^{(\text{gam})}(s)\right)\right) = h(\Omega)$ and $\Omega$ is noise term.

(b) is obtained by letting $Z = \log(T)$, and the definition of entropy.

(c) is derived noting $h(\log(T)) = h(Z) = -E\log f_Z(Z)$. Since $f_{Z|S}(Z|s) = f_{T|S}(t|s)|dt/dz| = f_{T|S}(t|s).t$, we have

$$-E\log f_{Z|S}(Z|s) = -E\left[\log\left(f_{T|S}(T|s).T\right)\big|s\right]. \tag{S11}$$

(d) is obtained from the definition of the entropy and the multiply property of logarithm. Replacing (S11) in (S9), and $I(\Lambda_{exc};T|s) = I\left(b_1^{(\text{gam})}(s)\lambda_{exc} + b_0^{(\text{gam})}(s);T|s\right)$, $J(F_\Lambda)$ is obtained as

$$J = h(T|s) + (\kappa - 1)E(\log(T)|s) - h(\Omega) - \alpha'' E[T|s] - \beta''. \tag{S12}$$

Hence, (S12) is a Lagrange multiplier equation of optimization problem in (S2-a). With some mathematical manipulation, which are skipped here for brevity, the solution of optimization problem (S2-a) is given by Gamma distribution[3]

$$f_{T|S}(t|s) = \frac{\beta^\kappa t^{\kappa-1} e^{-\beta t}}{\Gamma(\kappa)} u(t). \tag{S13}$$

where $\beta$ and $\kappa$ are shaping and scaling parameters of Gamma distribution of ISIs obtained from the constraints $E(t|s) = \beta/\kappa$ and $E(\log(T)|s) = \psi(\kappa) - \log(\beta)$.

## B. Appendix B

The optimal distribution for $\lambda_{exc}$, $f_{\Lambda_{exc}|S}(\lambda_{exc}|s)$, which maximizes the cost function of (S1), is given by

$$f_{\Lambda_{exc}|S}(\lambda_{exc}|s) = \beta^\kappa b_1^{(\text{gam})}(s) \frac{\Gamma\left(m_0^{(\text{gam})}(s)\right)}{\Gamma(\kappa)\Gamma\left(m_0^{(\text{gam})}(s) - \kappa\right)} \frac{\left(\lambda_{exc} b_1^{(\text{gam})}(s) - \beta + b_0^{(\text{gam})}(s)\right)^{m_0^{(\text{gam})}(s) - \kappa - 1}}{\left(\lambda_{exc} b_1^{(\text{gam})}(s) + b_0^{(\text{gam})}(s)\right)^{m_0^{(\text{gam})}(s)}} \times \\ u(\lambda_{exc} b_1^{(\text{gam})}(s) - \beta + b_0^{(\text{gam})}(s)) \tag{S14}$$

where $\Gamma(.)$ denotes the Gamma function.

**Proof.** By replacing the Gamma distribution in (24), we have

$$\int_0^\infty d\lambda_{exc} f_{\Lambda_{exc}}(\lambda_{exc}) \times$$

$$\frac{\left(b_1^{(gam)}(s)\lambda_{exc} + b_0^{(gam)}(s)\right)^{m_2^{(gam)}(s)\lambda_{exc}^2 + m_1^{(gam)}(s)\lambda_{exc} + m_0^{(gam)}(s)} t^{m_2^{(gam)}(s)\lambda_{exc}^2 + m_1^{(gam)}(s)\lambda_{exc} + m_0^{(gam)}(s)-1} e^{-\left(b_1^{(gam)}(s)\lambda_{exc} + b_0^{(gam)}(s)\right)t}}{\Gamma(m_2^{(gam)}(s)\lambda_{exc}^2 + m_1^{(gam)}(s)\lambda_{exc} + m_0^{(gam)}(s))} \quad (S15)$$

$$= \frac{\beta^\kappa t^{\kappa-1} e^{-\beta t} e^{b_0^{(gam)}(s)t}}{\Gamma(\kappa)}.$$

By changing variable $v = b_1^{(gam)}(s)\lambda_{exc}$, we have

$$\frac{1}{b_1^{(gam)}(s)} \int dv f_{\Lambda_{exc}}\left(\frac{v}{b_1^{(gam)}(s)}\right) \times$$

$$\frac{\left(v + b_0^{(gam)}(s)\right)^{m_2^{(gam)}(s)\left(\frac{v}{b_1^{(gam)}(s)}\right)^2 + m_1^{(gam)}(s)\left(\frac{v}{b_1^{(gam)}(s)}\right) + m_0^{(gam)}(s)} t^{m_2^{(gam)}(s)\left(\frac{v}{b_1^{(gam)}(s)}\right)^2 + m_1^{(gam)}(s)\left(\frac{v}{b_1^{(gam)}(s)}\right) + m_0^{(gam)}(s)-1} e^{-vt}}{\Gamma(m_2^{(gam)}(s)\left(\frac{v}{b_1^{(gam)}(s)}\right)^2 + m_1^{(gam)}(s)\left(\frac{v}{b_1^{(gam)}(s)}\right) + m_0^{(gam)}(s))} \quad (S16)$$

$$= \frac{\beta^\kappa t^{\kappa-1} e^{-\beta t} e^{b_0^{(gam)}(s)t}}{\Gamma(\kappa)} u(t).$$

By simplification, we have

$$\frac{1}{b_1^{(gam)}(s)} \int dv f_{\Lambda_{exc}}\left(\frac{v}{b_1^{(gam)}(s)}\right) \frac{\left(v + b_0^{(gam)}(s)\right)^{m_2^{(gam)'}(s)v^2 + m_1^{(gam)'}v + m_0^{(gam)}(s)} t^{m_2^{(gam)'}v^2 + m_1^{(gam)'}v + m_0^{(gam)}(s)-1} e^{-vt}}{\Gamma(m_2^{(gam)'}v^2 + m_1^{(gam)'}v + m_0^{(gam)}(s))} \quad (S17)$$

$$= \frac{\beta^\kappa t^{\kappa-1} e^{-\beta t} e^{b_0^{(gam)}(s)t}}{\Gamma(\kappa)} u(t),$$

where, $m_2^{(gam)'}(s) = m_2^{(gam)}(s) / \left(b_1^{(gam)}(s)\right)^2$, $m_1^{(gam)'}(s) = m_1^{(gam)}(s) / \left(b_1^{(gam)}(s)\right)^2$. The closed form solution for the above integral integration does not exist. As, we can observein Figure S8A, $m_2^{(gam)}(s)$ and $m_1^{(gam)}(s)$, for $s \geq 20\%$ is near to zero, and also as we can see in Figure 7A, for $s \geq 20\%$, $m^{(gam)}$ is $\lambda_{exc}$ independent function, hence we can write (S17) as

$$\frac{1}{b_1^{(gam)}(s)} \int dv f_{\Lambda_{exc}}\left(\frac{v}{b_1^{(gam)}(s)}\right) \frac{\left(v + b_0^{(gam)}(s)\right)^{m_0^{(gam)}(s)} t^{m_0^{(gam)}(s)-1} e^{-vt}}{\Gamma(m_0^{(gam)}(s))} = \frac{\beta^\kappa t^{\kappa-1} e^{-\beta t} e^{b_0^{(gam)}(s)t}}{\Gamma(\kappa)} u(t). \quad (S18)$$

Using definition of Laplace transformation, we have

$$\frac{1}{b_1^{(\text{gam})}(s)} \mathcal{L}\left( f_{\Lambda_{exc}}\left(\frac{v}{b_1^{(\text{gam})}(s)}\right) \frac{\left(v + b_0^{(\text{gam})}(s)\right)^{m_0^{(\text{gam})}(s)} t^{m_0^{(\text{gam})}(s)-1}}{\Gamma(m_0^{(\text{gam})}(s))} \right) = \frac{\beta^\kappa t^{\kappa-1} e^{-\beta t} e^{b_0^{(\text{gam})}(s)t}}{\Gamma(\kappa)} u(t). \quad (S19)$$

Using inverse Laplace transformation, we have

$$\frac{1}{b_1^{(\text{gam})}(s)} f_\Lambda\left(\frac{v}{b_1^{(\text{gam})}(s)}\right) \frac{\left(v + b_0^{(\text{gam})}(s)\right)^{m_0^{(\text{gam})}(s)}}{\Gamma(m_0^{(\text{gam})}(s))} = \beta^\kappa \left(v - \beta + m_0^{(\text{gam})}(s)\right)^{m_0^{(\text{gam})}(s)-\kappa-1} u(v - \beta + b_0^{(\text{gam})}(s)), \quad (S20)$$

hence, we have

$$f_{\Lambda_{exc}}\left(\frac{v}{b_1^{(\text{gam})}(s)}\right) = \beta^\kappa b_1^{(\text{gam})}(s) \frac{\Gamma(m_0^{(\text{gam})}(s))}{\Gamma(\kappa)\Gamma\left(m_0^{(\text{gam})}(s)-\kappa\right)} \frac{\left(v - \beta + b_0^{(\text{gam})}(s)\right)^{m_0^{(\text{gam})}(s)-\kappa-1}}{\left(v + b_0^{(\text{gam})}(s)\right)^{m_0^{(\text{gam})}(s)}} u(v - \beta + b_0^{(\text{gam})}(s)), \quad (S21)$$

by replacing $v = b_1^{(\text{gam})}(s)\lambda_{exc}$, we have

$$f_{\Lambda_{exc}}(\lambda_{exc}) =$$
$$\beta^\kappa b_1^{(\text{gam})}(s) \frac{\Gamma(m_0^{(\text{gam})}(s))}{\Gamma(\kappa)\Gamma\left(m_0^{(\text{gam})}(s)-\kappa\right)} \frac{\left(\lambda_{exc} b_1^{(\text{gam})}(s) - \beta + b_0^{(\text{gam})}(s)\right)^{m_0^{(\text{gam})}(s)-\kappa-1}}{\left(\lambda_{exc} b_1^{(\text{gam})}(s) + b_0^{(\text{gam})}(s)\right)^{m_0^{(\text{gam})}(s)}} u(\lambda_{exc} b_1^{(\text{gam})}(s) - \beta + b_0^{(\text{gam})}(s)). \quad (S22)$$

### C. Appendix C

Neuronal synchronization affects the noise term and hence enhances $I_{bpj}$.

**Proof.** By replacing $I(\Lambda_{exc};T|s)$ from (S5) in (27), we have

$$I = h(T|s) + (\kappa-1)E(\log(T)|s) - h\left(\log(T)\Big|\log\left(b_1^{(\text{gam})}(s)\Lambda_{exc} + b_0^{(\text{gam})}(s)\right),s\right) - C, \quad (S23)$$

Let $Z = \log(T)$, similar to (S5), the third term of the above equation is given by

$$h\left(\log(T)\Big|\log\left(b_1^{(\text{gam})}(s)\Lambda_{exc} + b_0^{(\text{gam})}(s)\right),s\right) = h(Z|\log\left(b_1^{(\text{gam})}(s)\Lambda_{exc} + b_0^{(\text{gam})}(s)\right),s)$$
$$= h\left(T\Big|\log\left(b_1^{(\text{gam})}(s)\Lambda_{exc} + b_0^{(\text{gam})}(s)\right),s\right) - E\left[T\Big|\log\left(b_1^{(\text{gam})}(s)\Lambda_{exc} + b_0^{(\text{gam})}(s)\right),s\right] \quad (S24)$$

Fitting a Gamma distribution to $f(T|\Lambda_{exc},S)$ in (19), (S23) is obtained as

$$I = h(T|s) + (\kappa - 1)E(\log(T)|s) - C$$
$$- \int_{\varsigma}^{\infty} f_{\Lambda_{exc}|S}(\lambda_{exc}|s) \left( \log\left( \left( \frac{1}{b^{(gam)}(\lambda_{exc},s)} \right) \Gamma\left(m^{(gam)}(s,\lambda_{exc})\right) \right) + \left(1 - m^{(gam)}(s,\lambda_{exc})\right) \psi\left(m^{(gam)}(s,\lambda_{exc})\right) \right.$$
$$\left. + m^{(gam)}(s,\lambda_{exc}) - \frac{m^{(gam)}(s,\lambda_{exc})}{b^{(gam)}(s,\lambda_{exc})} \right) d\lambda_{exc}, \quad \text{(S25)}$$

where, $\varsigma$ is minimum value of $\lambda_{exc}$ with nonzero PDF. First and second term of above equation depends on $f_{T|S}(t|s)$, which is obtained from optimization problem (S2-a) and given by (22). In (22), it can be seen that $f_{T|S}(t|s)$ is independent of $s$, hence first and second terms of (S25) are independent of $s$. Third term is constant and independent of $s$, However, fourth term is noise term and depends on $s$. Therefore, neuronal synchronization reduces effect of noise by minimizing noise term.